\documentclass[%
 reprint,
superscriptaddress,
 amsmath,amssymb,
pra,
]{revtex4-2}

\setcitestyle{super}

\usepackage{graphicx}%
\usepackage{dcolumn}%
\usepackage{bm}%

\usepackage{physics}
\usepackage{soul}
\usepackage{color}

\begin{document}

\preprint{APS/123-QED}

\newcommand{\activestate}{n}

\title{A Size-Consistent Multi-State Mapping Approach to Surface Hopping}

\author{Joseph E.\ Lawrence}
\email{joseph.lawrence@nyu.edu}
\affiliation{Department of Chemistry and Applied Biosciences, ETH Zurich, 8093 Zurich, Switzerland}
\affiliation{Simons Center for Computational Physical Chemistry, New York University, New York, NY 10003, USA}
\affiliation{Department of Chemistry, New York University, New York, NY 10003, USA}
\author{Jonathan R.\ Mannouch}
\email{jonathan.mannouch@mpsd.mpg.de}
\affiliation{Hamburg Center for Ultrafast Imaging, Universit\"at Hamburg and the Max Planck Institute for the Structure and Dynamics of Matter, Luruper Chaussee 149, 22761 Hamburg, Germany}
\author{Jeremy~O.\ Richardson}
\email{jeremy.richardson@phys.chem.ethz.ch}
\affiliation{Department of Chemistry and Applied Biosciences, ETH Zurich, 8093 Zurich, Switzerland}

\date{\today}%

\begin{abstract}
We develop a multi-state generalisation of the recently proposed mapping approach to surface hopping (MASH) for the simulation of electronically nonadiabatic dynamics. This new approach extends the original MASH method to be able to treat systems with more than two electronic states. It differs from previous approaches in that it is size consistent and rigorously recovers the original two-state MASH in appropriate limits. We demonstrate the accuracy of the method by application to a series of model systems for which exact benchmark results are available, and find that the method is well suited to the simulation of photochemical relaxation processes. 

\end{abstract}

\maketitle

\section{Introduction}
The Born-Oppenheimer approximation is foundational to the  study of modern molecular science and is highly accurate for the majority of systems at equilibrium. However, for many non-equilibrium processes, particularly those involving light--matter interaction, the approximation breaks down.\cite{Worth2004BeyondBO,Curchod2018review,Barbatti2011FSSHReview,Subotnik2016review,Gomez2020IntroChapterNonAdDyn,Mai2020FSSHChapter} Such processes are central to   a wide range of important fields including atmospheric chemistry, astrochemistry, as well as artificial and biological light harvesting. The ability to accurately simulate electronically nonadiabatic dynamics is therefore an essential tool in modern chemistry.
Unfortunately, without the Born-Oppenheimer approximation, molecular dynamics are significantly more complicated to simulate, and the development of new trajectory methods remains an area of active research.\cite{Stock2005nonadiabatic,Miller2016Faraday,Kelly2013SH,Martens2016CSH,Min2015nonadiabatic,Ha2018XFSH,MASH,spinmap,multispin,Esch2021EhrenfestTAB,Liu2021mapping,spinPLDM1}

The simplest approach to the simulation of nonadiabatic dynamics is Ehrenfest theory,\cite{Ehrenfest1927,Mittleman1961Ehrenfest,Delos1972Ehrenfest} where nuclei evolve under the mean-field potential of the current electronic state. While still very popular, the mean-field nature of the force can lead to unphysical nuclear dynamics particularly in molecular photochemistry where it fails to correctly describe wavepacket bifurcation. In 1990 Tully suggested an approach to overcome these issues,\cite{Tully1990hopping} called fewest switches surface hopping (FSSH), in which the nuclei always evolve on an adiabatic potential with stochastic hops between the surfaces determined by the electronic dynamics. FSSH has become the go-to method for simulating electronically nonadiabatic dynamics.\cite{Barbatti2011FSSHReview,Subotnik2016review,Gomez2020IntroChapterNonAdDyn,Mai2020FSSHChapter} Its popularity can be attributed to a number of key factors, including reasonable accuracy in describing simple photochemical problems and its ease of interpretation. Perhaps most importantly, however, is that it is highly economical with its use of information about the electronic structure of the system, making it well suited to ab-initio simulation.

However, FSSH is not without its own issues. A review of FSSH by Subotnik \emph{et al.}\cite{Subotnik2016review}~summarised its three key drawbacks as
\begin{enumerate}
    \item Overcoherence error: Although FSSH avoids the often unphysical mean-field force of Ehrenfest, in systems with multiple avoided crossings the wavefunction can become inconsistent with the active surface leading to errors.\cite{Schwartz1994ET,Bittner1995Decoherence,Schwartz1996Decoherence,Fang1999FSSH_inconsistency,Fang1999FSSH_inconsistency,Schwartz1996Decoherence,Wong2002Decoherence,Wong2002Decoherence2,Jasper2005Decoherence,Subotnik2011AFSSH,Jain2016AFSSH}
    \item Diabatic initialisation and measurement: It is unclear how to correctly initialise in or measure a diabatic population.\cite{Muller1997FSSH,Kelly2013SH,Schwerdtfeger2014ET,Landry2011hopping,Fuji2010Furan}
    \item Quantum Classical Liouville Equation (QCLE): The QCLE provides a rigorous framework for mixed quantum--classical simulations, however, FSSH is not rigorously derivable from the quantum classical Liouville equation (QCLE).\cite{Kapral1999MQCD,Shi2004QCLE,Bonella2010QCLE,Kelly2012mapping,Subotnik2013QCLE,Kapral2016FSSH}
\end{enumerate}
Each of these issues can in part or whole be attributed to the ad-hoc nature of the original FSSH algorithm, and hence ultimately to the final of the three drawbacks. A large effort has gone into deriving FSSH from the QCLE\@.\cite{Kapral1999MQCD,Shi2004QCLE,Bonella2010QCLE,Kelly2012mapping,Subotnik2013QCLE,Kapral2016FSSH} 
While a connection can be made, it requires rather restrictive assumptions.\cite{Subotnik2013QCLE} This perhaps explains why, although many suggestions have been made to address the drawbacks of FSSH, no one solution has become universally adopted. Without a rigorous derivation from the QCLE no modification can be unambiguously deemed `correct', leading to continued debate around the optimal algorithm. This ambiguity has also led to a variation in the details of the algorithm, including continued discussion over the correct way to treat frustrated hops,  \cite{HammesSchiffer1994FSSH,Muller1997FSSH,Sifain2016FSSH,Limbu2023RPSH} and a range of ways for performing momentum rescaling.\cite{Mai2018SHARC}

Recently a new surface hopping method has been proposed, the mapping approach to surface hopping (MASH).\cite{MASH}
Like FSSH, MASH is an independent classical trajectory method, in which the nuclei predominantly evolve on a single adiabatic surface with occasional hopping events between the surfaces. It is therefore of comparable computational cost to FSSH\@. Unlike FSSH, however, MASH is a rigorous short time approximation to the QCLE.\cite{MASH}
This has a number of benefits; not only is there a unique momentum rescaling and treatment of frustrated hops, but also a unique prescription for how to initialise and calculate diabatic populations. Furthermore, the MASH trajectories cannot become inconsistent in the way that they can in FSSH, and this has been shown to significantly reduce the effect of the overcoherence error.\cite{MASHrates,Molecular_Tully_JM}   
Finally, its connection to the QCLE allows MASH to be systematically improved via a ``quantum jump'' procedure,  and for rigorous decoherence corrections to be derived.\cite{MASH}

Unfortunately, the original MASH method was only derived in the special case of two electronic states.\cite{MASH} Recently, Runeson and Manolopoulos have proposed a version of MASH for multiple electronic states.\cite{Runeson2023MASH, Runeson2024MASH} However, their method does not recover the original two-state version of MASH\@. While this may not be an issue in certain systems, we shall argue that for the kind of photochemical problems we are interested in studying, their method is not an ideal solution to the problem of generalising MASH to multiple states.
It is therefore the aim of this paper to develop a generalisation of MASH to multiple states that recovers the original two-state theory. The resulting theory should not only recover the original theory for two-state problems, but should also be size consistent, so as to recover the original theory even when treating a multi-state problem in the limiting case that two states are uncoupled from the others.

\section{Background} \label{sec:background}
Before introducing our multi-state generalisation of MASH, we begin by reviewing some of the key ideas of the original two-state MASH approach and discussing the challenges with its generalisation to multiple states. One of the key practical differences between MASH and FSSH is that in MASH the active surface is obtained deterministically. In the two state case it is determined based on which wavefunction coefficient is largest, such that, labelling the upper adiabat $2$ and the lower adiabat $1$ the active state, $\activestate$, is defined as
\begin{equation}
    \activestate = h(|c_2|^2-|c_1|^2) + 1,
\end{equation}
where $h(x)$ is the Heaviside step function. This can more naturally be cast in terms of the Bloch sphere
\begin{subequations}
\begin{alignat}{3}
     S^{(2,1)}_x &= c_2c^*_1+c_2^*c_1 \\
     S^{(2,1)}_y &= i[c_2c^*_1-c_2^*c_1] \\
     S^{(2,1)}_z &= |c_2|^2-|c_1|^2.
\end{alignat}
\end{subequations}
The active surface is then defined according to which hemisphere of the Bloch sphere the system is in, with $\activestate = h(S^{(2,1)}_z) + 1$. One might be worried that, because the active surface is deterministic, MASH like Ehrenfest would be unable to describe wavepacket splitting. However, just as with other mapping approaches,\cite{spinmap} MASH overcomes this issue by representing a pure initial electronic state in terms of an ensemble over the Bloch sphere. For example, to measure the time-dependent population of a system which starts in the pure adiabatic state $2$, the ensemble is sampled from the distribution $\rho_2(S^{(2,1)}_z)=2 h(S^{(2,1)}_z) |S^{(2,1)}_z|$, with $S^{(2,1)}_x$ and $S^{(2,1)}_y$ chosen uniformly from the corresponding circle on the Bloch sphere. In this way MASH replaces the stochastic nature of the hops in FSSH with sampling over the initial conditions.

The challenge with generalising the MASH approach to more than two states stems from the fact that mapping approaches are inherently not size extensive. By this we mean that methods such as the Meyer--Miller--Stock--Thoss mapping (MMST)\cite{Meyer1979nonadiabatic,Stock1997mapping} and spin-mapping\cite{spinmap} do not give the same result if an uncoupled subsystem is treated separately or as part of a larger set of states.\cite{ellipsoid,thermalization,truncated} One of the most obvious effects of this is that it can lead to unphysical transitions between uncoupled states. This issue will also be present in any generalisation of MASH that evolves an initial distribution of wavefunction coefficients under the time-dependent electronic Schr\"odinger equation and determines the active state based on which state has the largest wavefunction coefficient. To some extent this error is reduced in MMST and spin-mapping by the non-positive definite nature of the statistics. In Runeson and Manolopoulos's version of MASH, they modified the method to be closer to other mapping approaches by introducing non-positive definite statistics and accepting the lack of size extensivity. To avoid ambiguity, and since their method is different to the original MASH even for two states, we will in the following refer to their method as a surface hopping inspired approach to mapping (SHIAM). In our multi-state generalisation of MASH, however, we would like to exactly recover the original two-state MASH and have a method that is size extensive with respect to the inclusion of additional uncoupled states.

\section{Theory: A multi-state generalisation of MASH}
\subsection{Dynamics}
In our proposed multi-state generalisation of MASH,
we introduce a separate effective Bloch sphere to describe the interaction between the active surface, $\activestate$, and each of the other $N-1$ electronic states, $\bm{S}^{(\activestate,b)}$. The spheres are defined such that for a given active surface all $N-1$ spheres satisfy the condition, $S_z^{(\activestate,b)}>0$.
Between hopping events we define the force on the nuclei to be given by the usual surface hopping expression
\begin{equation}
    \bm{F} = - \frac{\partial V_{\activestate}}{\partial \bm{q}}
\end{equation}
and the dynamics of each sphere to obey the equation
\begin{equation}
\label{eq:spin_dynamics}
    \hbar \dot{\bm{S}}^{(\activestate,b)} = \begin{pmatrix} 0 \\ \sum_k\frac{2\hbar}{m_k}d^{(\activestate,b)}_k(\bm{q}) p_k \\ V_\activestate(\bm{q})-V_b(\bm{q}) \end{pmatrix} \times {\bm{S}}^{(\activestate,b)},
\end{equation}
i.e.~evolving as if the two states $n$ and $b$ were treated as an isolated subsystem. For this reason we call our new method the uncoupled spheres MASH (unSMASH) approach.

Hopping events are a trivial generalisation of the two state case. A hopping event between the active state and another adiabatic state, $b$, occurs when $S_z^{(\activestate,b)}(t_{\rm hop})=0$. This hopping event can only be successful (i.e., result in a change of active state) provided that the kinetic energy along the nonadiabatic coupling vector (NACV) is greater than the energy gap between the new state, i.e.
\begin{equation}
     E_{\rm kin}^{(d)}=\frac{1}{2} \frac{(\tilde{\bm{p}} \cdot \tilde{\bm{d}})^2}{\tilde{\bm{d}}\cdot\tilde{\bm{d}}}   >V_b(\bm{q}) - V_n(\bm{q}) ,
\end{equation}
where $\tilde{p}_k=p_k/\sqrt{m_k}$ and $\tilde{d}_k=d^{(n,b)}_k/\sqrt{m_k}$ are the mass-weighted momentum and derivative coupling vectors respectively. In the event that the trajectory does not have sufficient energy to hop, it is ``forbidden'' and the component of the mass-weighted momentum along the derivative coupling vector is reversed
\begin{equation}
    \tilde{\bm{p}} \leftarrow  \tilde{\bm{p}} - 2 \tilde{\bm{d}}\,\frac{\tilde{\bm{p}}\cdot\tilde{\bm{d}}}{\tilde{\bm{d}}\cdot \tilde{\bm{d}}}.
\end{equation}
In the case that there is sufficient energy to hop the active state changes from $\activestate_{\rm i}=n$ to $\activestate_{\rm f}=b$ and the momentum along the derivative coupling vector is scaled so as to conserve energy
\begin{equation}
\tilde{\bm{p}} \leftarrow \tilde{\bm{p}} + \left(\sqrt{\frac{E_{\rm kin}^{(d)}+V_n(\bm{q}) - V_b(\bm{q})}{E_{\rm kin}^{(d)}}}-1\right)\tilde{\bm{d}}\,\frac{\tilde{\bm{p}}\cdot\tilde{\bm{d}}}{\tilde{\bm{d}}\cdot \tilde{\bm{d}}}.
\end{equation}
 After a successful hop, the spheres are relabelled according to the following rule
\begin{equation} \label{eq:relabel1}
    \bm{S}^{(\activestate_{\rm f},\mu)} \leftarrow\begin{cases} \bm{S}^{(\activestate_{\rm i},\mu)} & \text{ if } \mu\neq \activestate_{\rm i} \\
    \bm{S}^{(\activestate_{\rm f},\mu)} & \text{ if } \mu = \activestate_{\rm i} 
    \end{cases},
\end{equation}
where it is helpful to make use of the following identity, that follows trivially from the definition of the Bloch-sphere
\begin{equation} \label{eq:relabel2}
    \begin{pmatrix}
        S_x^{(b,a)} \\
        S_y^{(b,a)} \\
        S_z^{(b,a)} 
    \end{pmatrix} =
  \begin{pmatrix}
        S_x^{(a,b)} \\
        -S_y^{(a,b)} \\
        -S_z^{(a,b)} 
    \end{pmatrix}.
\end{equation}
Relabelling the spheres, rather than say resampling or using the relations in Appendix~\ref{app:S_to_c_convert}, is of course a choice. However we note that this definition of relabelling ensures that the dynamics is equivalent to the original MASH method in the two-state case, and is size extensive with respect to the addition of states that are entirely uncoupled from or have trivial crossings with the states of interest.

\begin{figure}[t]
    \centering
    \includegraphics[width=1.0\linewidth]{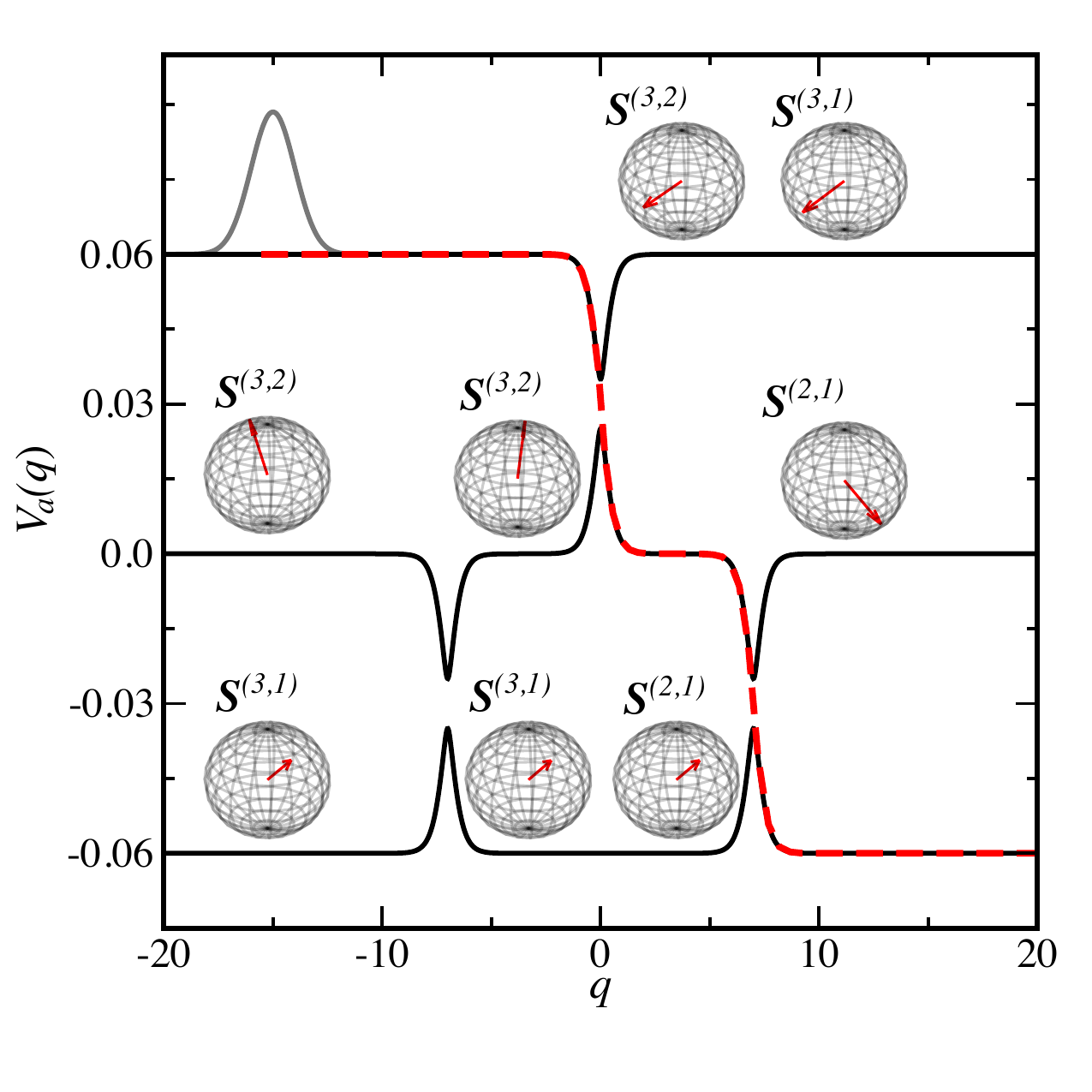}
    \caption{Diagram illustrating a typical unSMASH trajectory. Black lines show adiabatic potentials, the red line illustrates the path of a typical trajectory and the grey line illustrates the initial density in position space.  As there are three states there are always two effective Bloch spheres between the active surface and the other two states. The two Bloch spheres are shown at the corresponding values of $q$ along the trajectory, with their vertical positioning indicating the new surface that the system would hop to if the Bloch vector of that sphere switched hemisphere. After a successful nonadiabatic transition, the Bloch sphere not involved in the hop is then relabelled according to Eqs.~(\ref{eq:relabel1}) and (\ref{eq:relabel2}).}
    \label{fig:sphere_diagram}
\end{figure}

A schematic of a typical unSMASH trajectory involving two hops and the resulting relabelling of spheres is shown in Fig.~\ref{fig:sphere_diagram}.

\subsection{Initial Conditions and observables}

 Let us start by considering the definition of a time-dependent expectation value in the
Schr\"odinger representation
\begin{equation}
  \langle A(t) \rangle = \Tr[\hat{\rho}_t  \hat{A}]
\end{equation}
in which $\Tr[\dots]$ denotes a quantum trace over both the nuclear and electronic degrees of freedom. For a time-independent Hamiltonian the time-evolved density operator is given by
\begin{equation}
    \hat{\rho}_t = e^{-i\hat{H}t/\hbar} \, \hat{\rho} \, e^{+i\hat{H}t/\hbar}.
\end{equation} 
Here we choose to put $t$ as a subscript in order to highlight that we can consider $\hat{\rho}_t$ as a set of density operators labelled by the continuous variable, $t$. Taking a partial Wigner transform over the nuclear coordinates then gives
\begin{equation}
  \!\!\langle A(t) \rangle  =  \frac{1}{(2\pi\hbar)^f} \!\int\!\mathrm{d}\bm{q}_t \!\int\!\mathrm{d}\bm{p}_t \,\! \tr[\hat{W}_t(\bm{p}_t,\bm{q}_t)\hat{A}^{\rm w}(\bm{p}_t,\bm{q}_t)] \!\! \label{eq:Time_dependent_Expectation_val_Wigner}
\end{equation}
where $\tr[\dots]$ denotes a quantum trace over the electronic coordinates only, and we have chosen to label the integration variables by the time $t$ to emphasise that this is a phase space integral over a (pseudo) density at time $t$. The partial Wigner transform of the operator $\hat{A}$ is defined as
\begin{equation}
    \hat{A}^{\rm w}(\bm{p},\bm{q}) = \int\! \mathrm{d}\Delta\bm{q} \, e^{i \bm{p}\cdot\Delta \bm{q}/\hbar } \mel{\bm{q}+\frac{\Delta\bm{q}}{2}}{\hat{A}}{\bm{q}-\frac{\Delta\bm{q}}{2}}
\end{equation}
and we define the Wigner transformed density operator (Wigner distribution operator) as $\hat{W}_t(\bm{p},\bm{q})=\hat{\rho}^{\rm w}_t(\bm{p},\bm{q})$. 

\subsubsection{Adiabatic populations}
Before we consider how to treat general initial conditions and observables, we start by considering the simplest case where both $\hat{A}^{\rm w}(\bm{p},\bm{q})$ and $\hat{W}_0(\bm{p},\bm{q})$ contain only diagonal elements in the adiabatic basis 
\begin{equation}
    \hat{A}^{\rm w}(\bm{p},\bm{q}) = \sum_a {A}_{aa}^{\rm w} (\bm{p},\bm{q}) \dyad{a}{a}
\end{equation}
\begin{equation}
    \hat{W}_0(\bm{p},\bm{q}) = \sum_a {W}^{aa}_{0} (\bm{p},\bm{q}) \dyad{a}{a}.
\end{equation}
In this case the unSMASH approximation to the time-dependent expectation value is a trivial generalisation of the two-state case, and can be written as
\begin{equation}
\begin{aligned}
    &\langle A(t) \rangle  \approx \\
   &\tr_{\rm cl}\!\left[ \rho_{\rm P}(\mathbf{S}) \sum_{ab} W_0^{bb} (\bm{p},\bm{q}) P_b(\mathbf{S}) A^{\rm w}_{aa}\!\big(\bm{p}(t),\bm{q}(t)\big) P_a(\mathbf{S}(t))\right] 
   \end{aligned}
\end{equation}
where $\bm{p}(t)$, $\bm{q}(t)$ and $\mathbf{S}(t)$ are the unSMASH time evolved positions, momenta and effective Bloch spheres respectively. The projections onto adiabatic populations are replaced by $P_b(\mathbf{S})$ and $P_a(\mathbf{S}(t))$ which measure whether the system is in the corresponding active state e.g. 
\begin{equation}
    P_a(\mathbf{S}(t)) =  \delta_{a,\activestate(\mathbf{S}(t))}. \label{eq:MASH_projection_on_state_a}
\end{equation}
In this special case the initial density over the effective Bloch spheres is given by
\begin{equation}
    \rho_{\rm P}(\mathbf{S}) = \prod_{\mu\neq n}2|S_z^{(n,\mu)}|.
\end{equation}
To complete the specification we define the classical trace over the nuclear and electronic coordinates as
\begin{equation}
    \tr_{\rm cl}[\dots] = \frac{1}{(2\pi\hbar)^f} \int \mathrm{d}\bm{q} \int \mathrm{d}\bm{p} \int \mathrm{d}\mathbf{S} \, \dots
\end{equation}
where the integral over the effective Bloch spheres is given by
\begin{equation}
\begin{aligned}
    \int \mathrm{d}\mathbf{S}  &= \sum_{n=1}^N \prod_{\mu\neq n} \int \mathrm{d}\mathbf{S}^{(n,\mu)}  h(S_z^{(n,\mu)}) 
    \end{aligned} 
\end{equation}
in which each of the integrals over the individual effective Bloch spheres can be written in the form
\begin{equation}
    \int \mathrm{d}\mathbf{S}^{(n,\mu)}  = \frac{1}{2\pi} \int_0^{2\pi} \mathrm{d}\phi^{(n,\mu)} \int_0^{\pi} \mathrm{d}\theta^{(n,\mu)}\sin\theta^{(n,\mu)} 
\end{equation}
with
\begin{subequations}
\begin{equation}
     S_x^{(n,\mu)} = \sin\theta^{(n,\mu)}\cos\phi^{(n,\mu)}   
\end{equation}
\begin{equation}
     S_y^{(\mu)} = \sin\theta^{(n,\mu)}\sin\phi^{(n,\mu)}   
\end{equation}
\begin{equation}
     S_z^{(n,\mu)} = \cos\theta^{(n,\mu)}.   
\end{equation}
\end{subequations}

\subsubsection{Adiabatic coherences and diabatic populations}
In the general case the initial density and the observables may involve adiabatic coherences, for example for a system initialised in a diabatic population. To arrive at the final unSMASH expression for such systems one can trivially generalise the results from the original two-state MASH paper.\cite{MASH}  The unSMASH approximation to Eq.~(\ref{eq:Time_dependent_Expectation_val_Wigner}) is thus defined in terms of the unSMASH approximation to the time-dependent Wigner distribution operator,
\begin{equation}
\frac{\hat{W}_t(\bm{p}_t,\bm{q}_t)}{(2\pi\hbar)^f} \approx \tr_{\rm cl}[ \delta(\bm{p}_t-\bm{p}(t))\delta(\bm{q}_t-\bm{q}(t)) \hat{\omega}_t(\bm{p},\bm{q},\mathbf{S})]
\end{equation}
where the diagonal elements of $\hat{\omega}_t(\bm{p},\bm{q},\mathbf{S})$ are given by
\begin{equation}
    \mel{a_{\bm{q}(t)}}{\hat{\omega}_t(\bm{p},\bm{q},\mathbf{S})}{a_{\bm{q}(t)}}=  W^{\rm P}_0(\bm{p},\bm{q},\mathbf{S}) P_a(\mathbf{S}(t)) 
\end{equation}
and the off-diagonal elements ($a\neq b $) by
\begin{equation}
    \mel{a_{\bm{q}(t)}}{\hat{\omega}_t(\bm{p},\bm{q},\mathbf{S})}{b_{\bm{q}(t)}}=  W^{\rm C}_0(\bm{p},\bm{q},\mathbf{S}) \sigma_{ab}(\mathbf{S}(t))
\end{equation}
note that $\ket{a_{\bm{q}(t)}}$ indicates the adiabatic state at the unSMASH time evolved nuclear position $\bm{q}(t)$.
The measurement of the adiabatic population $P_a(\mathbf{S}(t))$ is defined in Eq.~(\ref{eq:MASH_projection_on_state_a}), and the coherence observables are defined as
\begin{equation}
   \! \sigma_{ab}(\mathbf{S}(t)) = \frac{S_x^{(a,b)}(t)-iS_y^{(a,b)}(t)}{2} \left(\delta_{\activestate(t),a}+ \delta_{\activestate(t),b} \right). \label{eq:unSMASH_off_diagonals}
\end{equation}
The final term, $W_0^{\rm X}(\bm{p},\bm{q},\mathbf{S})$, $\mathrm{X}=\mathrm{P}\text{ or }\mathrm{C}$, is the unSMASH transformed Wigner distribution operator defined as
\begin{equation}
    W_0^{\rm X}(\bm{p},\bm{q},\mathbf{S}) = \tr[\hat{w}_{\rm X}(\bm{q},\mathbf{S})\hat{W}_0(\bm{p},\bm{q})].
\end{equation}
where for populations we have
\begin{equation}
\begin{aligned}
     \hat{w}_{\rm P}(\bm{q},\mathbf{S}) &= \rho_{\mathrm P}(\mathbf{S})\dyad{\activestate(\mathbf{S})}{\activestate(\mathbf{S})} \\&+\sum_{a\neq \activestate(\mathbf{S})}\left( \dyad{\activestate(\mathbf{S})}{a} +\dyad{a}{\activestate(\mathbf{S})} \right) S_{x}^{(\activestate(\mathbf{S}),a)}\\
     &+\sum_{a\neq \activestate(\mathbf{S})}  i \left( \dyad{\activestate(\mathbf{S})}{a} -\dyad{a}{\activestate(\mathbf{S})} \right) S_{y}^{(\activestate(\mathbf{S}),a)}
\end{aligned}
\end{equation}
and for coherences we have
\begin{equation}
\begin{aligned}
     \hat{w}_{\rm C}(\bm{q},\mathbf{S}) &= 2 \dyad{\activestate(\mathbf{S})}{\activestate(\mathbf{S})} \\&+\frac{3}{2}\sum_{a\neq \activestate(\mathbf{S})}\left( \dyad{\activestate(\mathbf{S})}{a} +\dyad{a}{\activestate(\mathbf{S})} \right) S_{x}^{(\activestate(\mathbf{S}),a)}\\
     &+\frac{3}{2}\sum_{a\neq \activestate(\mathbf{S})}  i \left( \dyad{\activestate(\mathbf{S})}{a} -\dyad{a}{\activestate(\mathbf{S})} \right) S_{y}^{(\activestate(\mathbf{S}),a)}.
\end{aligned}
\end{equation}
Note to simplify notation the dependence of the adiabatic states on $\bm{q}$ has been suppressed, i.e.~$\ket{a}=\ket{a_{\bm{q}}}$.
These definitions guarantee that unSMASH recovers the original two-state MASH in the case that only two states are coupled. It therefore follows that unSMASH inherits the connection of MASH to the QCLE for systems in which only two states are coupled at a given time.

Given these definitions, the unSMASH observables for combinations of electronic operators can be obtained using simple matrix algebra.
For instance, to calculate a diabatic population $\hat{P}_j^{\rm dia}(t)=\dyad{j(t)}{j(t)}$
\begin{equation}
\begin{aligned}
    &\langle P_j^{\rm dia}(t) \rangle \approx \\
 &\tr_{\rm cl}\left[ \sum_{ab}\braket{j}{a_{\bm{q}(t)}} \!\!\! \mel{a_{\bm{q}(t)}}{\hat{\omega}_t(\bm{p},\bm{q},\mathbf{S})}{b_{\bm{q}(t)}} \!\! \braket{b_{\bm{q}(t)}}{j} \right]. \label{eq:diabatic_measurement} 
    \end{aligned}
\end{equation}
An example for how to compute $\hat{\omega}_t(\bm{p},\bm{q},\mathbf{S})$ for an initial diabatic density is given in Appendix~\ref{app:diabatic_populations}. 
\section{Results and Discussion}
In the following, in order to assess the accuracy of the unSMASH method, we consider a series of model systems for which exact results can be calculated for comparison.

In each case, the same number of trajectories (100000) and time step were used for both the FSSH and unSMASH calculations. Note the large number of trajectories were used to ensure that the differences between the methods was not due to statistical noise.
In practice, one can use far fewer trajectories to obtain reasonable results.
In fact, for adiabatic populations it is trivial to prove that unSMASH and FSSH require the same number of trajectories to achieve the same statistical convergence.

\subsection{Generalised Tully Model - Model X}
 The first system we consider is a three-state avoided crossing model (Model X) proposed by Subotnik in Ref.~\citenum{Subotnik2011MultiDimDecoherence}, for which the elements of the potential energy matrix in a diabatic basis can be written as
\begin{subequations}
\begin{align}
  V_{11}(q) &= A \left[\tanh(B q) + \tanh(B(q+7)) \right] \\
V_{22}(q) &= -A \left[\tanh(B q) + \tanh(B(q-7)) \right] \\
V_{33}(q) &= -A \left[\tanh(B (q+7)) - \tanh(B(q-7)) \right] \\
V_{12}(q) &= C \exp(- q^2) \\
V_{13}(q) &= C \exp(- (q+7)^2) \\
V_{23}(q) &= C \exp(- (q-7)^2),
\end{align}
\end{subequations}
where $A=0.03$, $B=1.6$ and $C=0.005$ with a mass of $m=2000$, all in atomic units.
This model can be considered as a three-state generalisation of a Tully I type model.\cite{Tully1990hopping} We consider here an initial Wigner density operator 
\begin{equation}
    \hat{W}_0(p,q) = 2\dyad{3_a}{3_a}  \exp(-\gamma(q-q_0)^2 - \frac{1}{\hbar^2\gamma}(p-p_0)^2)
\end{equation}
where $\ket{3_a}$ corresponds to the upper adiabatic state and the wavepacket starts on the left of all of the nonadiabatic crossings moving towards the right, $\gamma=1/2$, $q_0=-15$, $p_0=\sqrt{2mE_{\rm kin}}$ with $E_{\rm kin}=0.03$.  Note as the bulk of the distribution is far away from coupling regions this is equivalent to an initial wavepacket of the form
\begin{equation}
    \ket{\psi(q)} = \ket{3_a} \sqrt{\frac{\gamma}{\pi}} \exp(-\frac{\gamma}{2}(q-q_0)^2+ip_0q/\hbar). 
\end{equation}

\begin{figure}[t]
    \centering
    \includegraphics[width=1.0\linewidth]{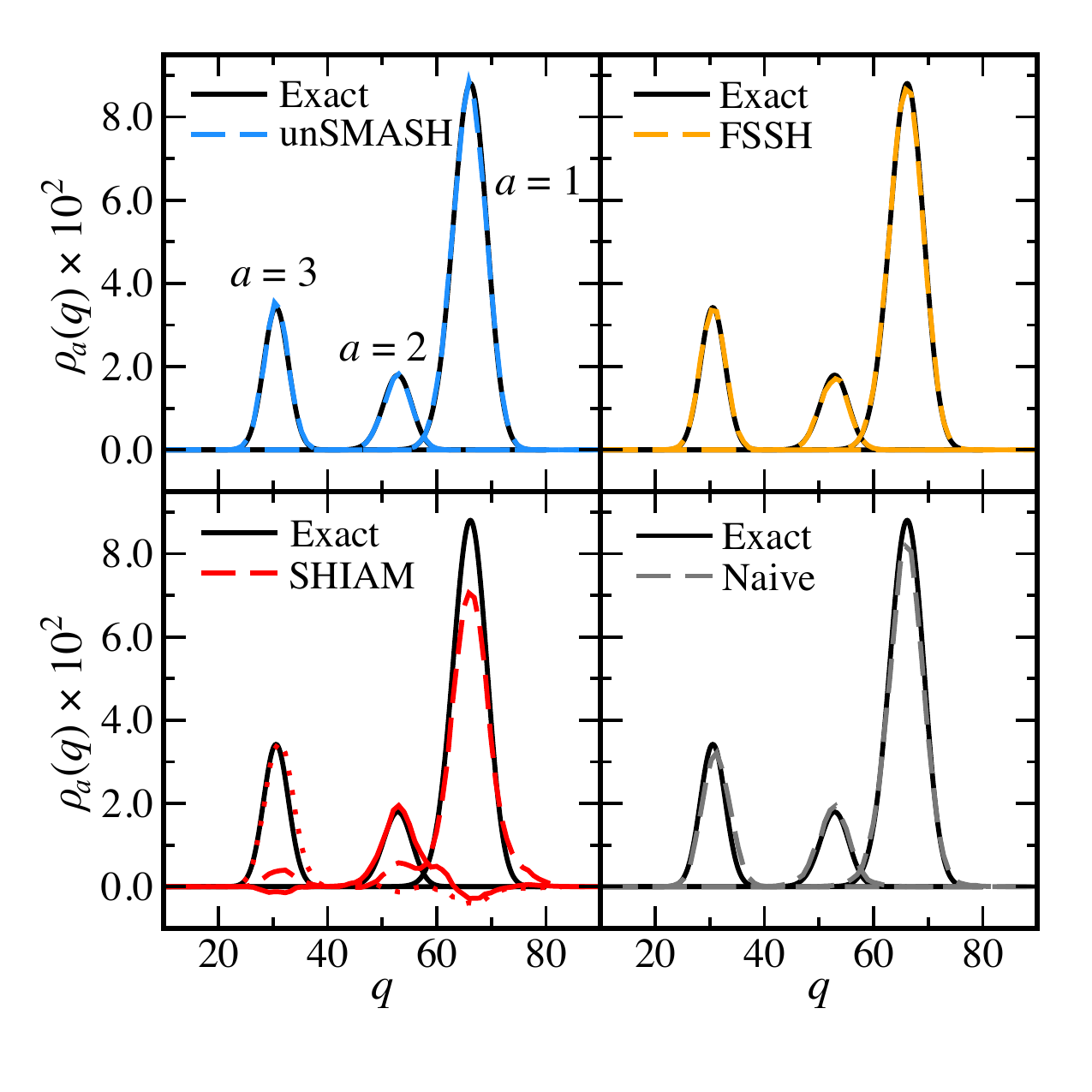}
    \caption{State dependent position distributions for Model X\cite{Subotnik2011MultiDimDecoherence}, at time $t=200\,$fs. For clarity the SHIAM results are given as a dotted, solid and dashed line for adiabatic states, 3, 2 and 1 respectively. (Note as mentioned in the main text SHIAM refers to the method of Runeson and Manolpoulos that they called ``multi-state MASH'' with cap initial conditions, renamed here to avoid confusion.)}
    \label{fig:One_D_model}
\end{figure}

Figure %
\ref{fig:sphere_diagram} depicts the model and initial position distribution.  Before considering the numerical results it is helpful to first consider the qualitative features of the model. As the initial wavepacket is on state $\ket{3_a}$ it should be essentially unaffected by the avoided crossing between states $\ket{2_a}$ and $\ket{1_a}$ at $q=-7$. Instead one expects that the wavepacket should stay on state $\ket{3_a}$ until it reaches the avoided crossing between $\ket{3_a}$ and $\ket{2_a}$ at $q=0$, at which point it should bifurcate, with part of the wavepacket accelerating and dropping onto state $\ket{2_a}$ and the remainder continuing on state $\ket{3_a}$. Finally the wavepacket on state $\ket{2_a}$ should bifurcate upon reaching the second avoided crossing between states $\ket{2_a}$ and $\ket{1_a}$, with essentially no effect to the wavepacket on state $\ket{3_a}$.
This therefore mimics the behaviour of a photochemically excited system that experiences a series of sequential avoided crossings. 
In order to probe the accuracy with which each bifurcation event is described we calculate the state-resolved nuclear density after the system has passed through all regions of nonadiabatic coupling, at $t=200\,$fs, which can be formally defined as
\begin{equation}
    \rho_a(q,t) = \big\langle \delta(\hat{q}(t)-q) \dyad{a(t)}{a(t)} \big\rangle .
\end{equation}
Exact results were calculated for comparison using a simple split-operator approach.\cite{Leforestier1991quantum}

In order to understand how the accuracy of unSMASH compares to other similar approaches, we also simulate the dynamics using FSSH, SHIAM and a naive multistate generalisation of MASH\@. In all three methods the momentum rescaling at frustrated and successful hops were treated in the same way as for unSMASH\@. For SHIAM the initial wavefunction coefficients were sampled using the cap method proposed in Ref.~\citenum{Runeson2023MASH}.
We have designed unSMASH to be size consistent with respect to  adding additional uncoupled electronic states, and to do so have introduced a modified electronic dynamics [Eq.~(\ref{eq:spin_dynamics})]. For comparison we therefore also consider a naive size-inconsistent generalisation of MASH\@. This naive approach differs from SHIAM as it retains the sampling and electronic observables used in the original two-state MASH, but is similar to SHIAM in that the wavefunction coefficients are evolved under the time-dependent Schr\"odinger equation and the active surface is determined as the state with the largest coefficient.\footnote{This is formally equivalent to the unSMASH prescription when one uses the definition in Appendix~\ref{app:S_to_c_convert} to convert back from the wavefunction coefficients to the normalised Bloch-spheres. This is the same as saying that, it is guaranteed that there will always be a unique state $n$ such that all $S_z^{(n,b)}$ are positive, and this state is the state with the largest $|c_n|^2$.}  When thinking in terms of wavefunction coefficients it is perhaps not immediately obvious how to generalise the original MASH sampling to multiple states. However, using the mapping between normalised two-state Bloch spheres and wavefunction coefficients described in Appendix~\ref{app:S_to_c_convert} we can simply use the same sampling as unSMASH for the sampling of the naive generalisation of MASH\@. As discussed in Sec.~\ref{sec:background} this method will be size inconsistent for the same reason as other mapping methods.

Figure \ref{fig:One_D_model} compares the predicted state dependent density $\rho_a(q,t)$ at $t=200\,$fs, where all trajectories have reached the product asymptotes, for all four methods against the exact result. Both MASH and FSSH are essentially in perfect agreement with the exact result. This indicates that both methods are capable of accurately capturing the correct wavepacket bifurcation at the two successive avoided crossings, and that they are not affected by any spurious transitions in regions of coupling between unpopulated states. 

In contrast, both SHIAM and the naive generalisation of MASH show clear differences when compared to the exact result. Considering first the naive generalisation of MASH (bottom right panel) one sees that the lack of size consistency results in errors in the branching ratios (peak areas) as well as shifts to the peaks, which can be attributed to spurious transitions between the surfaces that then result in unphysical forces being felt by the nuclei. 
In addition to these errors, the most obvious error in SHIAM is the presence of negative peaks in the probability distribution.  We note that this is not unique to three state systems and the same behavior is already observed in two state systems (as illustrated in Appendix \ref{app:Modified_Tully_1}). These negative peaks can be attributed to the non-positive definite statistics introduced by the definition of the electronic observables suggested by Runeson and Manolopoulos. This is unsurprising given similar negative probabilities are also observed in related mapping approaches such as MMST and spin-mapping. We stress, however, that SHIAM is not universally less accurate than the naive generalisation of MASH, or than unSMASH. In particular SHIAM has been shown to give very accurate results for the electronic populations in exciton models, such as those of the Fenna---Matthews---Olson complex, where the dynamics are in the fast nuclear (small reorganisation energy) limit.\cite{Runeson2023MASH,Runeson2024MASH} 
However, for this kind of system, involving a series of avoided crossings, it is clear that unSMASH is the preferred multi-state generalisation of MASH.

\subsection{Three-State Electron-Transfer Model}
One of the advantages of MASH is that it does not suffer from the same inconsistency (overcoherence) error as FSSH.\cite{MASH}
 This has been shown to lead to a significant improvement in the calculation of nonadiabatic rates in the limit of weak diabatic coupling,\cite{MASHrates}  i.e.~the Marcus Theory regime and of photodissociation product yields in molecular systems.\cite{Molecular_Tully_JM}
In order to test whether this advantage is retained in a multi-state system we consider a generalisation of the spin-boson model to three electronic states. 
We define the Hamiltonian in a diabatic basis, and decompose it into system and system-bath components 
\begin{equation}
    \hat{H} = \hat{H}_S +\hat{H}_{SB}.
\end{equation}
The system Hamiltonian is given (in mass-weighted coordinates) by
\begin{equation}
\begin{aligned}
    \hat{H}_S = &\frac{\hat{P}^2}{2} + \frac{1}{2}\Omega^2 \left(\hat{Q}+ \hat{\sigma}_z^{(0,2)}\frac{\kappa}{\Omega^2}\right)^2 \\+&\varepsilon\hat{\sigma}_z^{(0,2)}   + \Delta(\hat{\sigma}_x^{(0,1)}+\hat{\sigma}_x^{(1,2)})
    \end{aligned}
\end{equation}
with
\begin{subequations}
\begin{equation}
        \hat{\sigma}_z^{(i,j)} = \dyad{i}{i}-\dyad{j}{j}
\end{equation}
\begin{equation}
        \hat{\sigma}_x^{(i,j)} = \dyad{i}{j}+\dyad{j}{i}.
\end{equation}
\end{subequations}
Here $\hat{Q}$ can be thought of as a reaction coordinate along which the diabatic potentials corresponding to states $\ket{0}$, $\ket{1}$ and $\ket{2}$ are harmonic. The coupling between the diabatic states, mediated by $\Delta(\hat{\sigma}_x^{(0,1)}+\hat{\sigma}_x^{(1,2)})$, allows for transitions between the diabatic states in the vicinity of the crossings between diabats 0 \& 1, and diabats 1 \& 2. 
The system-bath part of the Hamiltonian describes the coupling of the reaction coordinate to a nuclear bath in the renormalised form
\begin{equation}
    \hat{H}_{SB} = \sum_{k=1}^{N_b}\frac{\hat{p}_k^2}{2} + \frac{1}{2}\omega_k^2 \left( \hat{q}_k + \frac{c_k \hat{Q}}{\omega_k^2} \right)^2.
\end{equation}
The effect of the bath on the system is encapsulated by the spectral density, which we take to be purely Ohmic
\begin{equation}
    J_{SB}(\omega) = \frac{\pi}{2}\sum_{k=1}^{N_b} \frac{c_k^2}{\omega_k} \delta(\omega-\omega_k) = \gamma \omega.
\end{equation}
This Hamiltonian acts as a simple model of sequential electron transfer from state 1 to 2 and then from state 2 to state 3.
Treating either of these pairs of states on their own gives a spin-boson model with a Marcus theory reorganisation energy defined as 
\begin{equation}
    \lambda = \frac{\kappa^2}{2\Omega^2}.
\end{equation}
Exact results can be calculated for this model using the hierarchical equations of motion (HEOM).\cite{Tanimura1989HEOM,Tanimura2020HEOM}  All HEOM calculations were performed using the HEOM-Lab code\cite{Fay2022LowTempHEOM,heomlab} and technical details are given in Appendix~\ref{app:HEOM_details}.

The parameters here are chosen to mimic the spin-boson model considered in Fig.~11 of Ref.~\citenum{MASH}. The parameters thus consist of a reorganisation energy, $\beta\lambda=1.5$, a low diabatic coupling, $\beta\Delta=0.25$, and a driving force {$\beta\varepsilon=2.5$} that corresponds to the Marcus inverted regime ($\varepsilon>\lambda$). The original model was a spin boson model with a Debye spectral density with $\beta\hbar\omega_c=1/20$, which corresponds here to $\beta\hbar\Omega^2/\gamma=1/20$ and $\Omega\gg\omega_c$. The frequency and friction were therefore taken to be $\beta\hbar\Omega=0.5$ and
$\gamma=10\Omega$. 
  We consider initial conditions corresponding to an initial diabatic population on state $\ket{0}$ with the nuclei in thermal equilibrium corresponding to the diabatic potential of state $\ket{1}$. The low frequency along the reaction coordinate ensures that nuclear quantum effects such as tunneling and zero-point energy are minimal, and allows us to take the classical limit of the initial thermal Wigner distribution. %
Additionally, for FSSH and unSMASH one can exactly integrate out the bath dynamics to give an additional Langevin friction and random force (with friction coefficient $\gamma$) along the reaction coordinate.\cite{Leggett1984spinboson,Garg1985spinboson,Thoss2001hybrid,Lawrence2019ET}  Having integrated out the bath, the initial reduced Wigner distribution operator is given by 
\begin{equation}
    \hat{W}_0(P,Q)=\beta\hbar\Omega \dyad{0}{0} e^{-\beta\left(\frac{P^2}{2} + \frac{1}{2}\Omega^2 Q^2  \right)}.
\end{equation}

In order to compare unSMASH to FSSH we need to use a version of FSSH where both the initial density and the observables are diabatic populations. A number of different ways of doing this have been proposed.\cite{Muller1997FSSH,Kelly2013SH,Schwerdtfeger2014ET,Landry2011hopping,Fuji2010Furan} Here we use the density matrix approach proposed by Landry \emph{et al.}~as this most closely resembles the way that diabatic initialisation and observables are treated within the MASH formalism. That they should be so similar is unsurprising given that Landry \emph{et al.}~based their approach on an analysis of the QCLE. Full details are given in Appendix~\ref{app:FSSH_density_matrix}. So as to compare to the most accurate version of FSSH we perform momentum rescaling at hops and frustrated hops in the same way as for MASH along the nonadiabatic coupling vector (which we label FSSH-nacv).
We note, however, that despite the long standing literature arguing that this should be the preferred treatment of momentum rescaling,\cite{Pechukas1969scattering2,Herman1984MomentumReversal} it is often not implemented in practice when using FSSH\@. 
For this reason, in addition to performing FSSH using the same momentum rescalings as MASH, we also consider here one of the most commonly used alternatives: rescaling along the velocity vector (which we label FSSH-vel).\cite{Plasser2014FSSH,Suchan2020LZSH,Papineau2024FSSH} This approach is popular because it can be used with electronic structure methods that do not provide access to the nonadiabatic coupling vector. However, it is not a size-consistent method (with respect to additional nuclear degrees of freedom). For this system--bath model, due to the formally infinite nature of the bath, it is in fact equivalent to performing no momentum rescaling at all. 

\begin{figure}[t]
    \centering
    \includegraphics[width=1.0\linewidth]{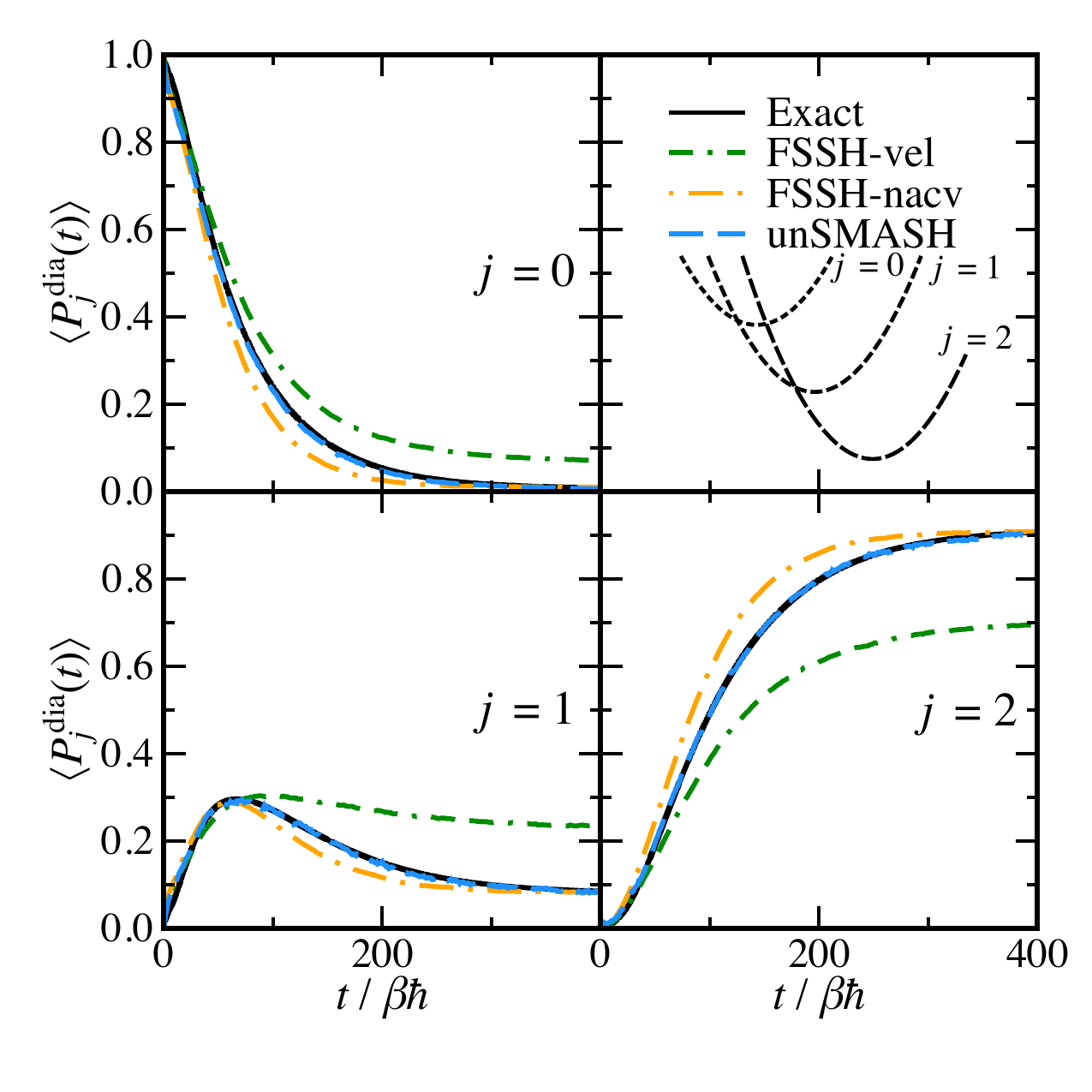}
    \caption{Diabatic populations for the three-state electron-transfer model illustrated in the top-right panel.
    The time step was 0.01\,$\beta\hbar$.
    }
    \label{fig:ET_model}
\end{figure}

Figure \ref{fig:ET_model} shows the population of each of the diabatic states calculated using unSMASH and FSSH compared to the exact results. 
One sees immediately that rescaling along the velocity vector leads to a total failure of FSSH  to correctly thermalise in the long time limit. 
While rescaling along the NACV improves the results at long time, the inconsistency (overcoherence) error leads to noticeable errors in the predicted timescale of the population transfer.
In contrast we see that unSMASH very accurately recovers the exact result as a function of time. 
This confirms that the unSMASH generalisation of two-state MASH retains the improved accuracy compared to FSSH observed in the analagous spin-boson model of Ref.~\citenum{MASH}. It thus follows that unSMASH is capable of treating a series of well separated nonadiabatic transitions. 
While the differences between FSSH-nacv and unSMASH in this system are relatively minor we note that as shown in Ref.~\citenum{MASHrates} these differences can become much more pronounced in systems with larger reorganisation energies, and correspondingly longer timescales. 

\subsection{Three State Benzene Cation and Pyrazine Models}
In order to test the applicability of unSMASH for photochemical problems we consider two classic examples of ultrafast molecular relaxation after photoexcitation: pyrazine and the benzene cation. In both cases there are existing three-state vibronic-coupling models that capture the key features of the relaxation dynamics. These models provide an excellent test system as despite their complexity it is possible to obtain exact benchmark quantum mechanical results. For both models the Hamiltonian can be written in the form
\begin{equation}
\begin{aligned}
    \hat{H} &= \sum_{k=0}^{N_{\rm n}-1}\left( \frac{1}{2}\omega_k \hat{p}_k^2 + \frac{1}{2}\omega_k \hat{q}_k^2\right) + \sum_{j=0}^2 E_j \dyad{j}{j} \\
    &+ \sum_{k=0}^{N_{\rm n}-1} \sum_{j=0}^2 \left( \kappa_{j,k} \hat{q}_k  + \Gamma_{j,k} \hat{q}^2_k  \right) \dyad{j}{j} \\&+ \sum_{k=0}^{N_{\rm n}-1} \sum_{j,j'=0}^2 \lambda_{j,j'}^{(k)} \hat{q}_k \dyad{j}{j'} \label{eq:VC_Hamiltonian}
    \end{aligned}
\end{equation}
where, $\ket{0},\ket{1},\ket{2}$ are the three diabatic states, $\hat{q}_k$ and $\hat{p}_k$ are the normal-mode coordinates and momenta respectively, $N_{\rm n}$ is the number of normal modes included in the model, and the values of the model parameters are given in Appendix \ref{app:params}\@. The initial Wigner distribution operators are given for both systems by
\begin{equation}
    \hat{W}_0(\bm{p},\bm{q}) = \dyad{j_{\rm i}}{j_{\rm i}}{(2\pi\hbar)^{N_{\rm n}}}  \exp(-\sum_{k=0}^{N_{\rm n}-1}  q_k^2 + p_k^2 )
\end{equation}
corresponding to vertical excitation of the ground state density onto the diabatic state $\ket{j_{\rm i}}$ according to the Franck--Condon principle.

In order to treat the diabatic initial conditions and observables we used the density matrix approach of Ref.~\citenum{Landry2013FSSH} for the FSSH calculations. Again, to provide a representative example of the results that can be expected from FSSH we consider two different versions of the FSSH algorithm. Firstly, to give what we believe to be the most accurate and well justified version of FSSH, we include results for FSSH-nacv, where momentum rescaling and frustrated hops are dealt with in the same manner as unSMASH, by considering the component of the momentum along the nonadiabatic coupling vector. Secondly, we consider a kind of velocity rescaling (FSSH-vel-$\infty$). Since both models we are considering have a significantly reduced dimension compared to the real system (24 down to 9 for pyrazine and 30 down to 5 for the benzene cation), to give a more accurate representation of the error introduced by using velocity rescaling in the full model we choose here to consider the limit of the velocity rescaling in the large-system limit, in analogy to the three-state electron-transfer model considered in the previous section. This means that FSSH-vel-$\infty$ corresponds to all hops being allowed.

\begin{figure}[t]
    \centering
    \includegraphics[width=1.0\linewidth]{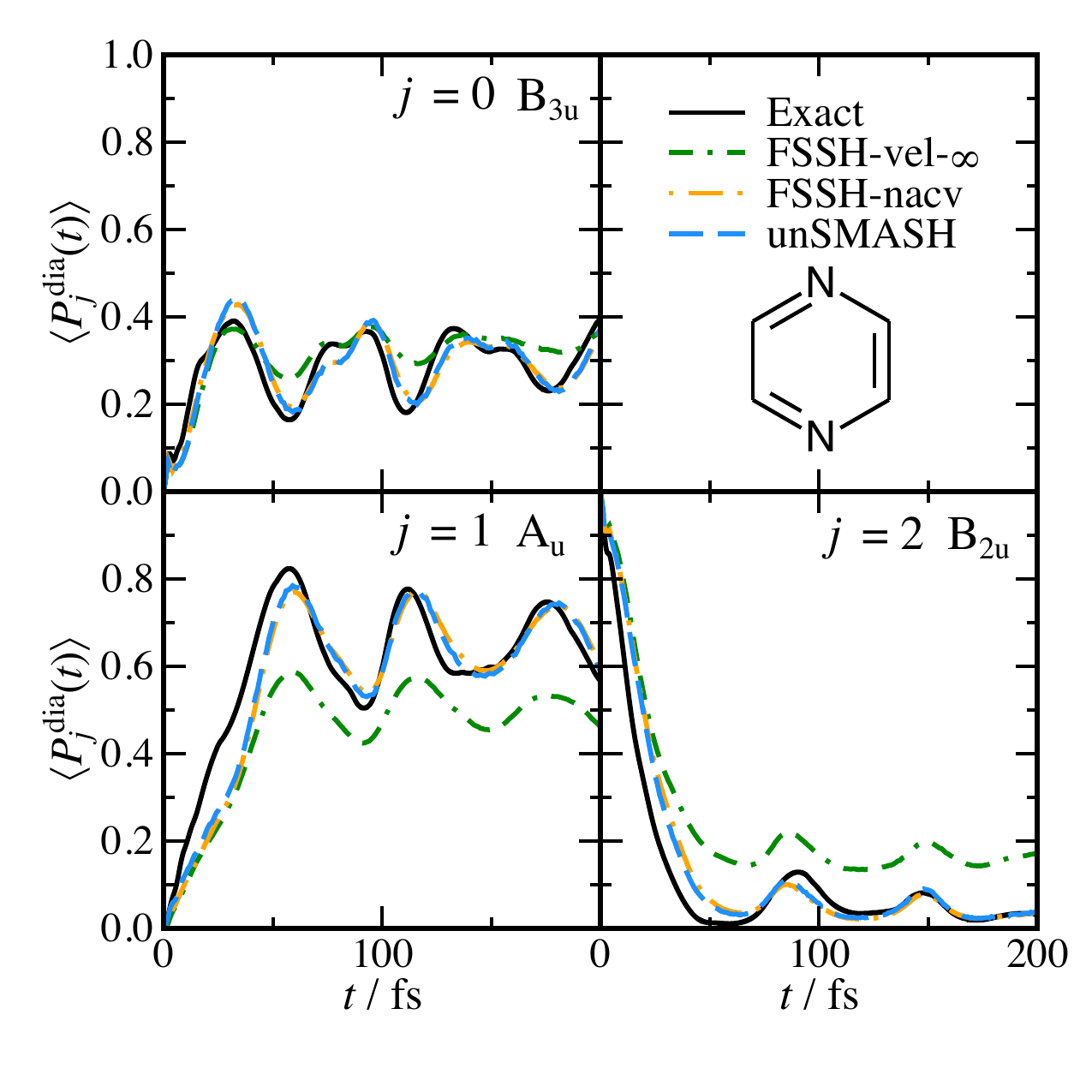}
    \caption{Diabatic populations as a function of time in the the three-state nine-mode model of pyrazine from Sala \emph{et al.}\cite{Sala2014PyrazineModel} Exact results are MCTDH calculations taken from Ref.~\citenum{Xie2019LZSH}.
    The time step for both FSSH and unSMASH calculations was 0.1\,fs.
    }
    \label{fig:Pyrazine}
\end{figure}
\begin{figure}[t]
    \centering
    \includegraphics[width=1.0\linewidth]{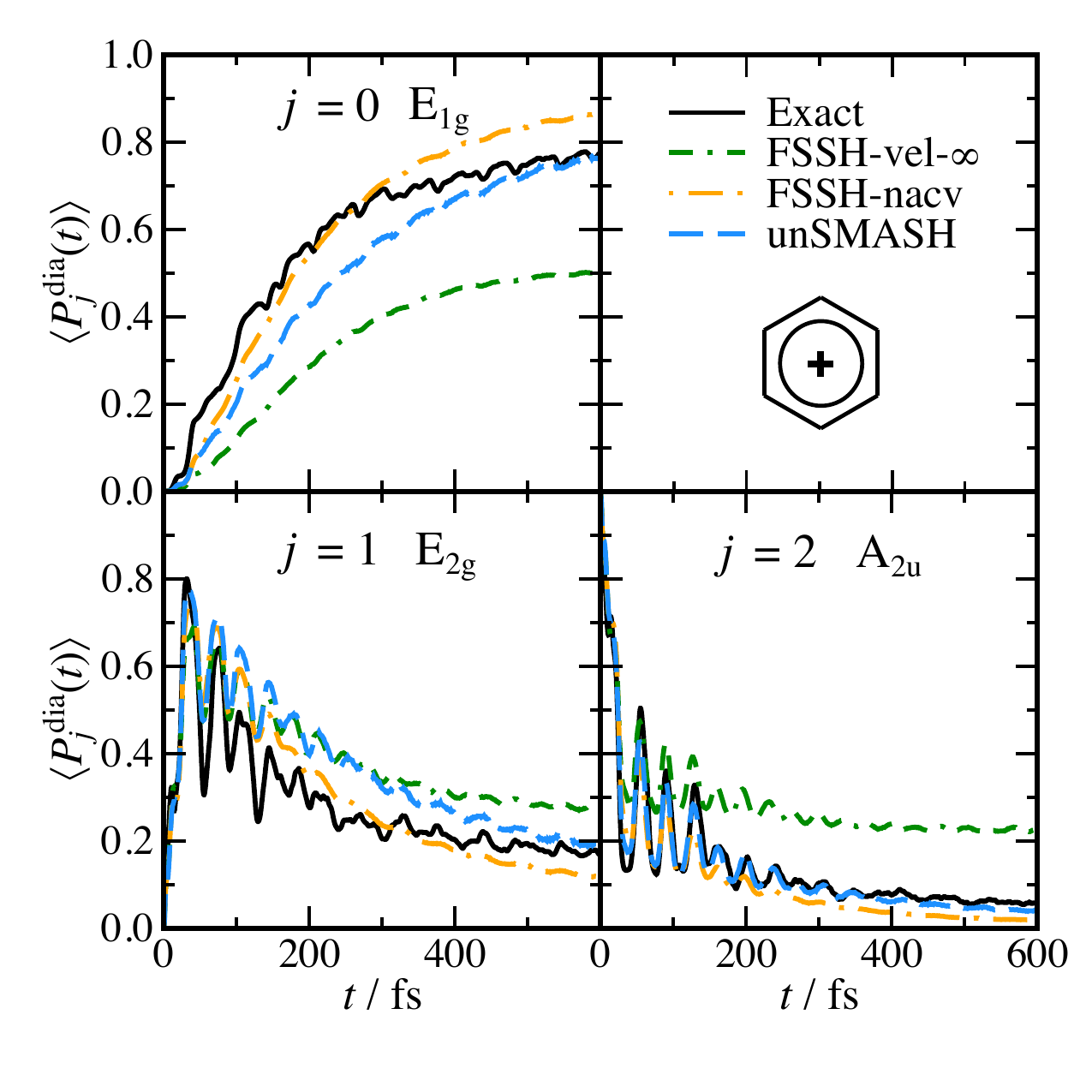}
    \caption{Diabatic populations as a function of time in the three-state five-mode model of the benzene cation from  K\"oppel.\cite{Koeppel1993benzene} Exact results calculated using DVR as described in Appendix \ref{app:DVR_details}.
    The time step for both FSSH and unSMASH calculations was 0.1\,fs.
    }
    \label{fig:Benzene_Cation}
\end{figure}

We begin by considering the ultrafast internal conversion of Pyrazine after excitation to $\mathrm{B}_{2\mathrm{u}}$ ($\ket{j_{\rm i}=2}$). The model we study here is the three-state nine-mode model of Sala \emph{et al.}~from Ref.~\citenum{Sala2014PyrazineModel}. For completeness we give the parameters in Appendix~\ref{app:Pyrazine_params}. The exact results for this model are provided by MCTDH calculations and were taken from Fig.~3 of Ref.~\citenum{Xie2019LZSH}. Figure \ref{fig:Pyrazine} shows the diabatic populations as a function of time. We again see that the velocity rescaling version of FSSH can lead to significant errors in the populations at long time. This is because using  velocity rescaling removes frustrated hops and so the system can unphysically return to the higher lying diabatic state, $\mathrm{B}_{2\mathrm{u}}$. The MASH derivation shows that a proper treatment of frustrated hops is required in order to correctly describe the nonadiabatic force term in the QCLE. It is therefore unsurprising that unSMASH and FSSH-nacv, which uses the same momentum rescaling scheme as unSMASH, are significantly more accurate that ${\text{FSSH-vel-}\infty}$. Both unSMASH and FSSH-nacv give very similar dynamics for this system. We see that both accurately describe the long-time populations for this system, as well as the qualitative features of the diabatic populations such as the period and magnitude of oscillations. The only notable difference between the trajectory simulations and the exact calculations is that the initial population transfer is slightly too slow.  It is interesting to note that in the 24-mode two-state model of pyrazine\cite{Raab1999Pyrazine} considered in the original MASH paper\cite{MASH} there was a much more significant difference between FSSH and MASH, with MASH being the more accurate of the two.

In the case of the Benzene cation we study the three-state five-mode model of K\"oppel,\cite{Koeppel1993benzene} considering relaxation after initial excitation to $\ket{j=2}$ which corresponds to the $\tilde{\text{C}}$ ($\mathrm{A}_{2\mathrm{u}}$) state.\cite{Koeppel1993benzene} For completeness we give the model parameters in Appendix~\ref{app:Benzene_Cation_Params}. Exact results were calculated using DVR, and full details are given in Appendix~\ref{app:DVR_details}. Figure \ref{fig:Benzene_Cation} compares the diabatic populations for all three diabatic states calculated using unSMASH and FSSH against the exact result. We see that this model exhibits somewhat different dynamics to that of pyrazine. After an initial ultra-fast transfer of population from $\ket{j=2}$ to $\ket{j=1}$ there is a slower population transfer to  $\ket{j=0}$. Again we see that while FSSH-vel-$\infty$ is accurate at very short time, the lack of frustrated hops leads to completely incorrect behaviour at long time. In contrast both unSMASH and FSSH-nacv capture the overall dynamics including the long time populations quite well.
The most notable differences between the exact result and those of FSSH-nacv and unSMASH are the populations of diabats $j=0$ and $j=1$ after about 50\,fs. The error clearly stems from a population transfer that is slightly too slow between these states with the population of $j=1$ consistently slightly too high and that of $j=0$ slightly too low. While similar, the results of unSMASH and FSSH-nacv do show some notable differences. Firstly considering the population in diabat $j=2$ we see that unSMASH captures the oscillations between 200 and 300\,fs more accuately than FSSH-nacv, as well as more accurately reproducing the long time limit. For the $j=0$ and $j=1$ states we see that while the timescales are similar between the two methods, FSSH-nacv predicts a slightly greater overall transfer of population from $j=1$ to $j=0$, resulting in a larger error at long time. It thus appears that while they are broadly similar, unSMASH is slightly more accurate than FSSH-nacv for this system. Importantly, however, both are significantly more accurate than FSSH-vel-$\infty$ emphasising the importance of the correct treatment of frustrated hops. This is perhaps unsurprising given we have chosen the features of FSSH-nacv to be as similar as possible to the MASH prescription (which is itself imposed by its connection to the QCLE). This highlights one of the important benefits of MASH: confirming best practice for FSSH simulations.

We note that the benzene cation was also studied in Ref.~\citenum{ultrafast} using various approaches including spin mapping.\cite{multispin} There it was found that spin mapping is better than FSSH at short time, and hence also better than unSMASH\@. However, it has been observed that going beyond a linear-vibronic model by adding even a small frequency difference between the diabatic states can significantly degrade the results of the spin-mapping method compared to MASH.\cite{MASH} It is nevertheless interesting to consider whether simple improvements to FSSH and unSMASH may be possible, for example by using a more accurate treatment of coherences or perhaps by explicit inclusion of nuclear quantum effects.

\section{Conclusion}
We have introduced a multi-state generalisation of the mapping approach to surface hopping (MASH). Our approach, which we call uncoupled spheres multi-state MASH (unSMASH), is size consistent with respect to adding additional uncoupled electronic states or nuclear degrees of freedom, and rigorously recovers the original two-state theory. It therefore inherits the connection of the original theory to the QCLE and is a rigorous short-time approximation under the assumption that only two states are coupled at a given time. We have demonstrated that it is as accurate or more accurate than FSSH for a series of model problems representative of typical photochemical systems.

Nevertheless there remain a number of interesting avenues for further research. 
One such area is the application of decoherence corrections. While 
it has been shown that MASH does not need decoherence corrections  as often as FSSH, and they have not been found to be necessary at all in any of the photochemical models we have studied, it may still be desirable to make use of them in certain situations.
One of the advantages of MASH is that its connection to the QCLE allows for the derivation of rigorous decoherence corrections. Given that unSMASH inherits these properties from the original theory, such decoherence corrections can also be rigorously applied to the present theory. At present these decoherence corrections are somewhat formal in nature, and a rigorous algorithm for determining when they should be applied is an interesting area of future research.  

While our present generalisation of MASH satisfies a number of necessary properties, there is one desirable property that it does not exhibit. That is that, it does not recover the exact electronic dynamics for a time-dependent Hamiltonian in a pure electronic system except in the two-state limit. This is important in systems where the nuclei are essentially unaffected by changes to the electronic state --- the fast nuclear limit. 
This means that unSMASH is not expected to work well for systems, such as typical models of the Fenna--Matthews--Olson complex,\cite{Tao2010FMO,Kim2012FMO,FMO,Runeson2023MASH} where the electronic states are all very close in energy and the potentials only slightly shifted with respect to one another.
Of course it is always possible to obtain the exact electronic dynamics in this limit by modifying the electronic observables, and this is what is done in SHIAM.\cite{Runeson2023MASH} However, one would ideally like the method to retain the connection between the electronic observables and the force on the nuclei as in the original two-level MASH\@. Without this one does not expect in general to obtain such accurate results in systems where the nuclei are more strongly coupled to the electronic coordinates, as the electronic observables can become inconsistent with the nuclear motion.  

A long-standing issue in the field is how best to accurately and practically incorporate nuclear quantum effects such as zero-point energy and tunneling in molecular simulations. As with FSSH and other methods based on classical trajectories, these effects are partially incorporated into MASH via the initial Wigner distribution. However, classical molecular dynamics doesn't obey quantum detailed balance,\cite{Hele2015Matsubara} which leads to the well-known phenomenon of zero-point energy leakage.\cite{Habershon2009water} For electronically adiabatic systems in the linear-response regime, this problem has been overcome by methods such as ring-polymer molecular dynamics (RPMD), that are based on imaginary-time path integrals, and for which the dynamics preserves the quantum Boltzmann distribution.\cite{RPMDcorrelation,Habershon2013RPMDreview,Lawrence2020rates} This has led to a number of electronically nonadiabatic extensions of RPMD being proposed, however, at present none is entirely satisfactory.\cite{Shushkov2012RPSH,mapping,Kretchmer2016KCRPMD,Duke2015MVRPMD,Chowdhury2017CSRPMD,Tao2019RPSH,Lawrence2019isoRPMD} There has however been recent development in imaginary-time path-integral methods capable of describing electronically nonadiabatic rates,\cite{Lawrence2018Wolynes,inverted,thiophosgene,Lawrence2020FeIIFeIII,Lawrence2020Improved}  and it might be hoped that such methods could be extended to develop a fully dynamical theory. 

While there may exist many interesting avenues for further theoretical development, we stress that unSMASH is already ready to be applied to real chemical problems. 
As an independent trajectory approach, it is well suited to on-the-fly ab-initio simulations and is no more expensive than FSSH\@.
Because of the similarity of the algorithms, many of the tricks developed for a robust implementation of FSSH, such as using wavefunction overlaps rather than nonadiabatic coupling vectors,\cite{Meek2014FSSHIntegrator,Jain2016AFSSH,Mai2020FSSHChapter,Jain2022FSSH} can be immediately picked up by MASH\@.
It is therefore in an immediate position to be applied
to study ab initio photochemical relaxation. In fact, we have already used the original two-state MASH to perform ab-initio simulations on a series of benchmark photochemical systems in conjunction with on-the-fly electronic structure theory at the levels of CASSCF and LR-TDDFT,\cite{Molecular_Tully_JM} and in a concurrent publication we have applied unSMASH to predict the relaxation of cyclobutanone after excitation to S$_2$ at the CASSCF level of theory.\cite{Lawrence2024cyclobutanone}

Here, we have demonstrated that unSMASH is at least as accurate as FSSH for describing typical ultrafast photochemical relaxation, and has a number of added advantages. 
It is more accurate for rate problems,\cite{MASHrates,Molecular_Tully_JM} and hence is capable of describing systems which exhibit ultrafast relaxation followed by a slower nonadiabatic process such as an electron transfer. Additionally, as MASH can is a short-time approximation to the QCLE there are a number of formal advantages. Firstly, there is no ambiguity in how momentum should be rescaled or what should be done at frustrated hops. Secondly, the connection to the QCLE gives a rigorous prescription for how to initialise or measure a diabatic population.  
These formal connections between MASH and the QCLE thus also help to confirm best practice for FSSH simulations, and we have reiterated that the accuracy of FSSH can be highly dependent on these details.  
Perhaps most exciting of all is that we do not believe this to be the final multi-state MASH method, it seems that there are still many further improvements around the corner.

\section*{Supplementary Material}
Exact quantum-mechanical benchmark data for the diabatic and adiabatic populations of the benzene cation.
\begin{acknowledgments}
JEL was supported by an Independent Postdoctoral Fellowship at the Simons Center for Computational Physical Chemistry, under a grant from the Simons Foundation (839534, MT) and JRM was supported by the Cluster of Excellence ``CUI: Advanced Imaging of Matter'' of the Deutsche Forschungsgemeinschaft (DFG) – EXC 2056 – project ID 390715994.
\end{acknowledgments}
\section*{Author Declarations}
\subsection*{Conflict of interest}
The authors have no conflicts to disclose.
\subsection*{Author Contributions}

J.E.L.~Conceptualization (equal), Investigation (lead), Methodology (equal), Software (lead), Validation (lead), Visualization (lead), Writing - original draft (lead), Writing - review \& editing (equal).
J.R.M.~Conceptualization (equal), Investigation (supporting), Methodology (equal),  Visualization (supporting), Writing - review \& editing (equal).
J.O.R.~Conceptualization (equal), Investigation (supporting), Methodology (equal), Software (supporting), Validation (supporting), Visualization (supporting), Writing - review \& editing (equal).

\appendix

\section{Diabatic initialisation and measurement} \label{app:diabatic_populations}
If the initial density involves a diabatic population
\begin{equation}
    \hat{W}_0(\bm{p},\bm{q}) = W_{\rm n}(\bm{p},\bm{q})\dyad{j}{j},
\end{equation}
where $W_{\rm n}(\bm{p},\bm{q})/(2\pi\hbar)^f$ is the distribution of nuclear degrees of freedom,
then we can evaluate the trace in the definition of $W_0^{\rm P}$ to give
\begin{equation}
    \begin{aligned}
     W_0^{\rm P}(\bm{p},\bm{q},\mathbf{S}) &= W_{\rm n}(\bm{p},\bm{q})\Big(  \rho_{\mathrm P}(\mathbf{S})|\braket{j}{\activestate(\mathbf{S})}|^2 \\&+\sum_{a\neq \activestate(\mathbf{S})} 2\Re\Big(\braket{j}{\activestate(\mathbf{S})}\braket{a}{j}\Big) S_{x}^{(\activestate(\mathbf{S}),a)}\\
     &-\sum_{a\neq \activestate(\mathbf{S})}  2\Im\Big(\braket{j}{\activestate(\mathbf{S})}\braket{a}{j}\Big) S_{y}^{(\activestate(\mathbf{S}),a)} \Big) \\
     &=W_{\rm n}(\bm{p},\bm{q}) g_j^{\rm P}(\bm{q},\mathbf{S})
\end{aligned}
\end{equation}
and the definition of $W_0^{\rm C}$ to give
\begin{equation}
    \begin{aligned}
     W_0^{\rm C}(\bm{p},\bm{q},\mathbf{S}) &= W_{\rm n}(\bm{p},\bm{q})\Big(  2|\braket{j}{\activestate(\mathbf{S})}|^2 \\&+\sum_{a\neq \activestate(\mathbf{S})} 3\Re\Big(\braket{j}{\activestate(\mathbf{S})}\braket{a}{j}\Big) S_{x}^{(\activestate(\mathbf{S}),a)}\\
     &-\sum_{a\neq \activestate(\mathbf{S})}  3\Im\Big(\braket{j}{\activestate(\mathbf{S})}\braket{a}{j}\Big) S_{y}^{(\activestate(\mathbf{S}),a)} \Big) \\
     &=W_{\rm n}(\bm{p},\bm{q}) g_j^{\rm C}(\bm{q},\mathbf{S}).
\end{aligned}
\end{equation}
Inserting these results into Eq.~(\ref{eq:diabatic_measurement}) we can then obtain a simple expression for evaluating diabatic populations at time $t=0$ for a system initially in a diabatic population
\begin{equation}
\begin{aligned}
    &\langle P_j^{\rm dia}(t) \rangle \approx \\
 &\phantom{+}\tr_{\rm cl}\left[ \sum_{a}\braket{j}{a_{\bm{q}(t)}} \!\!\! \mel{a_{\bm{q}(t)}}{\hat{\omega}_t(\bm{p},\bm{q},\mathbf{S})}{a_{\bm{q}(t)}} \!\! \braket{a_{\bm{q}(t)}}{j} \right]
 \\
 &+\tr_{\rm cl}\left[ \sum_{a\neq b}\braket{j}{a_{\bm{q}(t)}} \!\!\! \mel{a_{\bm{q}(t)}}{\hat{\omega}_t(\bm{p},\bm{q},\mathbf{S})}{b_{\bm{q}(t)}} \!\! \braket{b_{\bm{q}(t)}}{j} \right]\\
 &=\tr_{\rm cl}\left[ \sum_{a}\left|\!\braket{j}{a_{\bm{q}(t)}}\! \right|^2 W_{\rm n}(\bm{p},\bm{q}) g_j^{\rm P}(\bm{q},\mathbf{S}) P_a(\mathbf{S}(t)) \right]
 \\
 &+\tr_{\rm cl}\left[ \sum_{a\neq b}\braket{j}{a_{\bm{q}(t)}} \!\! \braket{b_{\bm{q}(t)}}{j} W_{\rm n}(\bm{p},\bm{q}) g_j^{\rm C}(\bm{q},\mathbf{S}) \sigma_{ab}(\mathbf{S}(t)) \right]\\
 &=\left\langle N\sum_{a}\left|\!\braket{j}{a_{\bm{q}(t)}}\! \right|^2  g_j^{\rm P}(\bm{q},\mathbf{S}) P_a(\mathbf{S}(t)) \right\rangle
 \\
 &+\left\langle N \sum_{a\neq b}\braket{j}{a_{\bm{q}(t)}} \!\! \braket{b_{\bm{q}(t)}}{j}  g_j^{\rm C}(\bm{q},\mathbf{S}) \sigma_{ab}(\mathbf{S}(t)) \right\rangle, 
    \end{aligned}
\end{equation}
where in the final line we have introduced the expectation value taken over the distribution
\begin{equation}
    \rho(\bm{p},\bm{q},\mathbf{S}) = \frac{1}{N(2\pi\hbar)^f} W_{\rm n}(\bm{p},\bm{q}),
\end{equation}
which corresponds to an equal probability of being in any initial active state with the nuclei sampled from the distribution $W_{\rm n}(\bm{p},\bm{q})/(2\pi\hbar)^f$. It is of course trivial to conceive of more efficient sampling schemes, for example by weighting each state by $|\langle j | \activestate(\mathbf{S})\rangle|^2$, however we will leave further discussion of this to future publications.

\section{Converting from normalised two-state Bloch spheres to wavefunction coefficients} \label{app:S_to_c_convert}
We give here the key equations for converting from the set of $N-1$ normalised Bloch spheres $\{\bm{S}^{(n,b)}; b=1,\dots,N \text{ and } b\neq n\}$, defined by Eq.~(\ref{eq:normalised_bloch_spheres}), back to the wavefunction coefficients. We begin by noting that the normalised Bloch spheres contain $2(N-1)$ independent variables, and hence do not store information about the normalisation of the wavefunction or the overall phase. We are therefore free to define the wavefunction to be normalised, and the overall phase such that $c_n$ is real and positive. Given this a simple rearrangement of the definition of $S_z^{(n,b)}$ allows one to show that
\begin{equation}
    \frac{|c_b|^2}{|c_n|^2} = \frac{1-S_z^{(n,b)}}{1+S_z^{(n,b)}}
\end{equation}
and hence that
\begin{equation}
   c_n = \sqrt{\frac{1}{\sum_{b\neq n}\frac{1-S_z^{(n,b)}}{1+S_z^{(n,b)}}+1}}
\end{equation}
and  
\begin{equation}
    |c_b| = c_n \sqrt{\frac{1-S_z^{(n,b)}}{1+S_z^{(n,b)}}}.
\end{equation}
Now all that remains is to determine the phases, $c_b=|c_b|e^{i\phi_b}$. This can be done by noting that
\begin{subequations}
    \begin{equation}
        S_x^{(n,b)} = \frac{c_n |c_b|}{|c_n|^2+|c_b|^2} 2\cos(\phi_b)
    \end{equation}
    \begin{equation}
        S_y^{(n,b)} = \frac{c_n |c_b|}{|c_n|^2+|c_b|^2} 2\sin(\phi_b)
    \end{equation}
\end{subequations}
such that 
\begin{equation}
  \phi_b = \arg\!\left(\frac{S_x^{(n,b)}+i S_y^{(n,b)}}{2} \frac{|c_n|^2+|c_b|^2}{c_n |c_b|}\right).
\end{equation}

\section{Negative peaks in SHIAM results for two state Tully model}
\label{app:Modified_Tully_1}
\begin{figure}[t]
    \centering
    \includegraphics[width=1.0\linewidth]{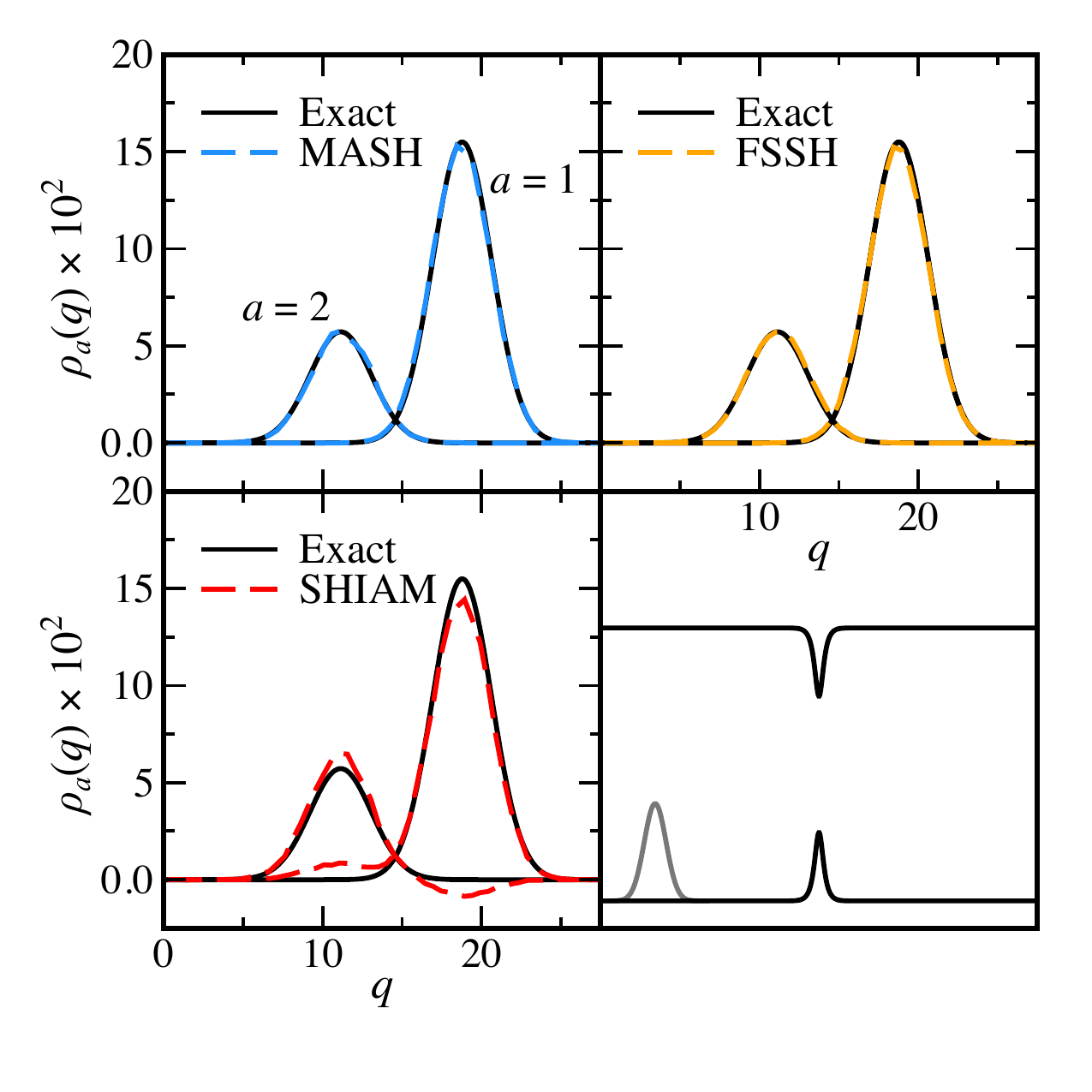}
    \caption{State-dependent position distributions at $t=150$\,fs for the modified Tully model I  of Ref.~\citenum{Ananth2007SCIVR}. The initial conditions are chosen to be equivalent ``Low Energy'' results depicted in Fig.~5 of Ref.~\citenum{MASH} and Fig.~6 of Ref.~\citenum{Runeson2023MASH}. Bottom rate panel shows illustration of model potentials and initial conditions. (Note as mentioned in the main text SHIAM refers to the method of Runeson and Manolpoulos that they called ``multi-state MASH'' with cap initial conditions, renamed here to avoid confusion.)}
    \label{fig:Tully_Model_I}
\end{figure}

Here, for completeness we illustrate that the negative peaks observed in the SHIAM predictions of the Model X position distribution are not a feature of the multi-state model considered, but are already present in systems with only two states. This is illustrated in Fig.~\ref{fig:Tully_Model_I} which shows the probability distribution at $t=150$\,fs for the modified Tully model I  of Ref.~\citenum{Ananth2007SCIVR}, with initial conditions corresponding to the ``Low Energy'' results of Fig.~5 of Ref.~\citenum{MASH} and Fig.~6 of Ref.~\citenum{Runeson2023MASH}. We see that while MASH and FSSH both faithfully reproduce the exact results, SHIAM shows spurious extra peaks (positive and negative) in the region associated with the other state. 

\section{FSSH Diabatic initial conditions and observables}
\label{app:FSSH_density_matrix}
In order to provide the fairest comparison between MASH and FSSH we make use of the density-matrix approach to FSSH\cite{Landry2013FSSH} for the calculation of diabatic properties. In the following we describe the approach and point out its similarities to and differences from the unSMASH method. Firstly, as with the unSMASH approach the density-matrix approximation to the time-evolved Wigner distribution operator can be written in the form 
\begin{equation}
\frac{\hat{W}_t(\bm{p}_t,\bm{q}_t)}{(2\pi\hbar)^f} \approx \tr_{\rm cl}[ \delta(\bm{p}_t-\bm{p}(t))\delta(\bm{q}_t-\bm{q}(t)) \hat{\omega}^{\rm FSSH}_t(\bm{p},\bm{q},\bm{c})].
\end{equation}
One of the major differences here is that the classical trace is now defined as
\begin{equation}
    \tr_{\rm cl}[\dots] = \frac{1}{(2\pi\hbar)^f}\int \mathrm{d}\bm{q}\int \mathrm{d}\bm{p}\int \mathrm{d}\bm{c}\, \dots
\end{equation}
where the integral over the space of wavefunction coefficients is given by\cite{Runeson2023MASH}
\begin{equation}
   \int \mathrm{d}\bm{c} \, \dots = \frac{N!}{(2\pi)^N} \prod_{a=1}^N \int \mathrm{d} c^{\rm Re}_a \int \mathrm{d} c^{\rm Im}_a \,\delta(r-1) \, \dots
\end{equation}
with $r=\sqrt{\sum_a |c_a|^2}$ being the norm of the wavefunction. Here the local density operator is given by a similar expression to that for unSMASH, with the diagonal elements given by
\begin{equation}
    \!\mel{a_{\bm{q}(t)}}{\hat{\omega}^{\rm FSSH}_t(\bm{p},\bm{q},\bm{c})}{a_{\bm{q}(t)}}=  W^{\rm FSSH}_0(\bm{p},\bm{q},\bm{c}) \delta_{n(t),a} 
\end{equation}
and the off-diagonal elements by
\begin{equation}
    \!\mel{a_{\bm{q}(t)}}{\hat{\omega}^{\rm FSSH}_t(\bm{p},\bm{q},\bm{c})}{b_{\bm{q}(t)}}=  W^{\rm FSSH}_0(\bm{p},\bm{q},\bm{c}) \sigma_{ab}(t) ,
\end{equation}
where the off-diagonal measurement is calculated as
\begin{equation}
\sigma_{ab}(t)=\frac{c_a(t) c_b^*(t)}{|c_a(t)|^2+|c_b(t)|^2} \left(\delta_{\activestate(t),a}+ \delta_{\activestate(t),b} \right).
\end{equation}
It is interesting to note that this is can be written in the same form as Eq.~(\ref{eq:unSMASH_off_diagonals}) by defining the (normalised) FSSH effective Bloch spheres as
\begin{subequations}
\label{eq:normalised_bloch_spheres}
    \begin{equation}
        S_x^{(a,b)} = \frac{c_a^{*}c_b+c_a c_b^{*}}{|c_a(t)|^2+|c_b(t)|^2}
    \end{equation}
    \begin{equation}
        S_y^{(a,b)} = \frac{-ic_a^{*}c_b+ic_a c_b^{*}}{|c_a(t)|^2+|c_b(t)|^2}
    \end{equation}
    \begin{equation}
        S_z^{(a,b)} = \frac{|c_a|^2-|c_b|^2}{|c_a(t)|^2+|c_b(t)|^2} .
    \end{equation}
\end{subequations}
The key difference however is that the FSSH initial density, $W_0^{\rm FSSH}(\bm{p},\bm{q},\bm{c})$, cannot be written in terms of a Weyl kernel. Instead, we first define a $(\bm{p},\bm{q})$ dependent (normalised) basis that diagonalises the initial density,
\begin{equation}
    \hat{W}_0(\bm{p},\bm{q})=\sum_{\alpha}{W}^{\alpha\alpha}_0(\bm{p},\bm{q}) \dyad{\alpha(\bm{p},\bm{q})}{\alpha(\bm{p},\bm{q})}
\end{equation}
then the FSSH initial density can be written as
\begin{equation}
    W^{\rm FSSH}_0(\bm{p},\bm{q},\bm{c}) = \sum_{\alpha} {W}^{\alpha\alpha}_0(\bm{p},\bm{q}) \delta_{\bm{c}}(\bm{c}-\bm{c}^\alpha(\bm{p},\bm{q})) ,
\end{equation}
where the elements of $\bm{c}^\alpha(\bm{p},\bm{q})$ are given by
\begin{equation}
    {c}^\alpha_a(\bm{p},\bm{q}) = \langle a | \alpha(\bm{p},\bm{q})\rangle
\end{equation}
and as a small technical detail the multidimensional delta function $\delta_{\bm{c}}(\bm{c}-\bm{c}^\alpha(\bm{p},\bm{q}))$ is defined such that it behaves as one would intuitively expect over the integration domain 
\begin{equation}
    \int \mathrm{d}\bm{c} \, f(\bm{c}) \delta_{\bm{c}}(\bm{c}-\bm{c}^\alpha(\bm{p},\bm{q})) = f(\bm{c}^{\alpha}(\bm{p},\bm{q})).
\end{equation}

\section{HEOM technical details}
\label{app:HEOM_details}
For the purpose of the HEOM calculations it is computationally advantageous to take the reaction coordinate out of the system degrees of freedom and treat it as part of the bath, by performing a normal mode transformation. The Hamiltonian can then be written in the form
\begin{equation}
    \hat{H} = \hat{H}_e + \hat{H}_n + \hat{V}_{en}
\end{equation}
where the electronic part of the Hamiltonian is given by
\begin{equation}
    \hat{H}_e = \varepsilon\hat{\sigma}_z^{(0,2)} -\lambda \dyad{1}{1} + \Delta(\hat{\sigma}_x^{(0,1)}+\hat{\sigma}_x^{(1,2)})
\end{equation}
and the nuclear part by
\begin{equation}
    \hat{H}_{n} = \sum_{k=0}^{N_b} \frac{\hat{\tilde{p}}_k}{2} +\frac{1}{2}\tilde{\omega}_k \hat{\tilde{q}}_k^2.
\end{equation}
The influence of the nuclear bath on the electronic dynamics is then completely specified by the electron nuclear coupling 
\begin{equation}
    \hat{V}_{en} = \hat{\sigma}_z^{(0,2)}\sum_{k=0}^{N_b} \tilde{c}_k \hat{\tilde{q}}_k
\end{equation}
and the spectral density
\begin{equation}
   J_{en}(\omega) = \frac{\pi}{2}\sum_{k=0}^{N_b} \frac{\tilde{c}_k^2}{\tilde{\omega}_k} \delta(\omega-\tilde{\omega}_k) =  \frac{\kappa^2 \gamma \omega}{(\omega^2-\Omega^2)^2+\gamma^2\omega^2}.
\end{equation}
This spectral density, that results from taking the normal mode transformation,\cite{Garg1985spinboson,Leggett1984spinboson,Thoss2001hybrid} is known as the Brownian oscillator spectral density and is straightforward to treat using HEOM.\cite{Tanimura1989HEOM,Tanimura2020HEOM}

\section{Model Parameters}
\label{app:params}
\subsection{Pyrazine}
\label{app:Pyrazine_params}
For completeness here we give the parameters of the three state Pyrazine model of Ref.~\citenum{Sala2014PyrazineModel}. The vibrational energies in each mode are
\begin{equation*}
\begin{aligned}
                &\omega_0   = 0.073495 \text{ eV} \text{  mode $\nu_{6\mathrm{a}}$ ($\mathrm{a}_\mathrm{g}$)} \\
                &\omega_1   = 0.126150 \text{ eV} \text{ mode $\nu_1$ ($\mathrm{a}_\mathrm{g}$)} \\
                &\omega_2   = 0.153991 \text{ eV} \text{ mode $\nu_{9\mathrm{a}}$ ($\mathrm{a}_\mathrm{g}$)} \\
                &\omega_3   = 0.199006 \text{ eV} \text{ mode $\nu_{8\mathrm{a}}$ ($\mathrm{a}_\mathrm{g}$)} \\
                &\omega_4   = 0.115999 \text{ eV} \text{  mode $\nu_{10\mathrm{a}}$ ($\mathrm{b}_{1\mathrm{g}}$)} \\
                &\omega_5   = 0.090953 \text{ eV} \text{  mode $\nu_4$ ($\mathrm{b}_{2\mathrm{g}}$)} \\
                &\omega_6   = 0.116741 \text{ eV} \text{  mode $\nu_5$ ($\mathrm{b}_{2\mathrm{g}}$)} \\
                &\omega_7   = 0.167660 \text{ eV} \text{ mode $\nu_3$ ($\mathrm{b}_{3\mathrm{g}}$)} \\
                &\omega_8   = 0.192537 \text{ eV} \text{ mode $\nu_{8\mathrm{b}}$ ($\mathrm{b}_{3\mathrm{g}}$)} \\
\end{aligned}
\end{equation*}
The diabatic potentials at $\bm{q}=0$ are given by
\begin{equation*}
\begin{aligned}
                &E_0 = 3.931201 \text{ eV} \text{ $B_{3u}$ }(n\to\pi^*) \\
                &E_1 = 4.450000 \text{ eV} \text{ $A_u$  }(n\to\pi^*) \\
                &E_2 = 4.791332 \text{ eV} \text{ $B_{2u}$ }(\pi\to\pi^*)
\end{aligned}
\end{equation*}
and the linear intra-state couplings by
\begin{equation*}
\begin{aligned}
                &\kappa_{0, 0} = -0.081046 \text{ eV }\\
                &\kappa_{1, 0} = -0.167811 \text{ eV }\\
                &\kappa_{2, 0} =  0.127832 \text{ eV }\\
                &\kappa_{0, 1} = -0.038299 \text{ eV }\\
                &\kappa_{1, 1} = -0.083091 \text{ eV }\\
                &\kappa_{2, 1} = -0.183131 \text{ eV }\\
                &\kappa_{0, 2} =  0.117396 \text{ eV }\\
                &\kappa_{1, 2} = -0.070680 \text{ eV }\\
                &\kappa_{2, 2} =  0.045362 \text{ eV }\\
                &\kappa_{0, 3} = -0.086844 \text{ eV }\\
                &\kappa_{1, 3} = -0.465185 \text{ eV }\\
                &\kappa_{2, 3} =  0.026224 \text{ eV }\\
\end{aligned}
\end{equation*}
with all other $\kappa_{n,j}=0$. The quadratic intra-state couplings are given by
\begin{equation*}
\begin{aligned}
                &\Gamma_{0, 4} = -0.012429 \text{ eV}\\
                &\Gamma_{1, 4} = -0.047533 \text{ eV}\\
                &\Gamma_{2, 4} = -0.012429 \text{ eV}\\
                &\Gamma_{0, 5} = -0.029919 \text{ eV}\\
                &\Gamma_{1, 5} = -0.030508 \text{ eV}\\
                &\Gamma_{2, 5} = -0.030508 \text{ eV}\\
                &\Gamma_{0, 6} = -0.014038 \text{ eV}\\
                &\Gamma_{1, 6} = -0.026064 \text{ eV}\\
                &\Gamma_{2, 6} = -0.026064 \text{ eV}\\
                &\Gamma_{0, 7} = -0.006172 \text{ eV}\\
                &\Gamma_{1, 7} = -0.006172 \text{ eV}\\
                &\Gamma_{2, 7} =  0.000631 \text{ eV}\\
                &\Gamma_{0, 8} = -0.011511 \text{ eV}\\
                &\Gamma_{1, 8} = -0.011511 \text{ eV}\\
                &\Gamma_{2, 8} =  0.007448 \text{ eV}\\
\end{aligned}
\end{equation*}
with all other $\Gamma_{n, j}=0$. Finally, the linear inter-state coupling constants are given by
\begin{equation*}
\begin{aligned}
                &\lambda_{0, 2}^{(4)} = \lambda_{2, 0}^{(4)} = 0.195323 \text{ eV }\\
                &\lambda_{1, 2}^{(5)} = \lambda_{2, 1}^{(5)} = 0.060269 \text{ eV }\\
                &\lambda_{1, 2}^{(6)} = \lambda_{2, 1}^{(6)} = 0.053232 \text{ eV }\\
                &\lambda_{0, 1}^{(7)} = \lambda_{1, 0}^{(7)} = 0.064514 \text{ eV }\\
                &\lambda_{0, 1}^{(8)} = \lambda_{1, 0}^{(8)} = 0.219400 \text{ eV }\\
\end{aligned}
\end{equation*}
with all other $\lambda_{j,j'}^{(k)}=0$.
\subsection{Benzene Cation}
\label{app:Benzene_Cation_Params}
Here for completeness we give the values of the parameters in the model Hamiltonian (Eq.~(\ref{eq:VC_Hamiltonian})) for the Benzene cation model of K\"oppel.\cite{Koeppel1993benzene} Note that there are some sign differences between the parameters given here and those reported in Ref.~\citenum{Koeppel1993benzene}, as these changes were found to be necessary to reproduce the reported quantum results of the original paper.\cite{JohanPhD}
The vibrational energies of each mode are given by
\begin{equation*}
\begin{aligned}
                &\omega_0 = 0.123 \text{ eV}
                \text{  mode $\nu_2$ }
                (\mathrm{a}_{1\mathrm{g}})\\
                &\omega_1 = 0.198 \text{ eV}
                \text{  mode $\nu_{16}$ }
                (\mathrm{e}_{2\mathrm{g}})\\
                &\omega_2 = 0.075 \text{ eV}
                \text{  mode $\nu_{18}$ }
                (\mathrm{e}_{2\mathrm{g}})\\
                &\omega_3 = 0.088 \text{ eV}
                \text{  mode $\nu_8$ }
                (\mathrm{b}_{2\mathrm{g}})\\
                &\omega_4 = 0.120 \text{ eV}
                \text{  mode $\nu_{19}$ }
                (\mathrm{e}_{2\mathrm{u}})
                \\
\end{aligned}
\end{equation*}
The diabatic potentials at $\bm{q}=0$ are given by
\begin{equation*}
\begin{aligned}
                &E_0 = 0.00 \text{ eV }\,\,\tilde{\mathrm{X}}\ (^2\mathrm{E}_{1\mathrm{g}}) \\
                &E_1 = 2.09 \text{ eV }\,\,\tilde{\mathrm{B}}\ (^2\mathrm{E}_{2\mathrm{g}})\\
                &E_2 = 2.69 \text{ eV }\,\, \tilde{\mathrm{C}}\ (^2\mathrm{A}_{2\mathrm{u}})\\
\end{aligned}
\end{equation*}
and the linear intra-state couplings by
\begin{equation*}
\begin{aligned}
                &\kappa_{0, 0} = -0.042 \text{ eV}\\
                &\kappa_{0, 1} = -0.246 \text{ eV}\\ 
                &\kappa_{0, 2} = -0.125 \text{ eV}\\ 
                &\kappa_{1, 0} = -0.042 \text{ eV}\\
                &\kappa_{1, 1} = \phantom{-} 0.242 \text{ eV}\\
                &\kappa_{1, 2} = \phantom{-} 0.1   \text{ eV}\\
                &\kappa_{2, 0} = -0.301 \text{ eV}\\
\end{aligned}
\end{equation*}
Note that the couplings $\kappa_{0, 1}$ and $\kappa_{0, 2}$ have opposite sign to those reported in Ref.~\citenum{Koeppel1993benzene}, and all other $\kappa_{j,k}=0$. For this model all quadratic intra-state couplings are zero. Finally, the linear inter-state coupling constants are given by
\begin{equation*}
    \begin{aligned}
        &\lambda_{0, 1}^{(3)} =\lambda_{1, 0}^{(3)} =0.164 \text{ eV}\\
        &\lambda_{1, 2}^{(4)} = \lambda_{2, 1}^{(4)} =0.154 \text{ eV}\\
    \end{aligned}
\end{equation*}
with all other $\lambda_{j, j'}^{(k)}=0$.

\section{Exact quantum-mechanical benchmarks for benzene cation}
\label{app:DVR_details}
The vibronic-coupling model of the benzene cation was simulated by wavepacket dynamics on a five-dimensional grid using the discrete-variable representation (DVR). \cite{Light1985DVR} %
In many aspects we follow the calculation details from K\"oppel, \cite{Koeppel1993benzene} who found convergence up to 200~fs using 18, 26, 28, 14, 12 basis functions for modes $\nu_2$, $\nu_{16}$, $\nu_{18}$, $\nu_8$, $\nu_{19}$, respectively.
Note that our Hermite DVR is equivalent to K\"oppel's finite-basis representation using harmonic-oscillator eigenstates.
As we wished to propagate up to 600~fs, a larger basis of 22, 30, 32, 18, 16 was used.
We used a short iterative Lanczos propagator \cite{Leforestier1991quantum} with a Krylov subspace of order 7 and a timestep of 0.25~fs.
All operations on the wavefunction were implemented on an NVIDIA Tesla GPU (graphical processing unit) using CuPy. \cite{cupy_learningsys2017}
The propagation was carried out in the diabatic representation and adiabatic populations were obtained using the eigenvectors of the diabatic potential energy matrix at each DVR point.
Whereas K\"oppel reports his calculations taking 50 hours on a CPU (central processing unit) in the 1990s, thanks to improvements in modern computing, we are able to obtain results in about 15 minutes.

\bibliography{references,extra_refs}%

\end{document}